\documentclass[
11pt,
letterpaper,
reprint,
notitlepage,
superscriptaddress,
aps,prx
]{revtex4-2}

\usepackage{amsmath,amssymb,amsfonts} 

\usepackage{algorithm}
\usepackage{algpseudocode}

\algblock{Input}{EndInput}
\algnotext{EndInput}
\algblock{Output}{EndOutput}
\algnotext{EndOutput}

\usepackage{amsthm} 

\usepackage{empheq}
\usepackage[notrig]{physics}

\usepackage[normalem]{ulem}

\usepackage{mathrsfs}
\DeclareMathAlphabet\mathbfcal{OMS}{cmsy}{b}{n}

\usepackage{microtype}

\usepackage[svgnames,dvipsnames]{xcolor}
\usepackage{graphicx}
\definecolor{NewBlue}{rgb}{0.1, 0.1, 0.7}
\definecolor{NewRed}{rgb}{0.7, 0.1, 0.1}
\usepackage[colorlinks,
linkcolor=Maroon,
citecolor=NewBlue,
urlcolor=ForestGreen]{hyperref}
\usepackage[capitalise]{cleveref}

\newtheorem{lemma}{Lemma}
\newtheorem{theorem}{Theorem}
\newtheorem{definition}{Definition}

\crefname{algocf}{alg.}{algs.}
\Crefname{algocf}{Algorithm}{Algorithms}

\usepackage{xr}
\renewcommand{\t}[1]{\mathrm{{#1}}}

\newcommand{\avg}{\expval}

\usepackage{empheq}
\newcommand{\eqdef}{\coloneqq}
\newcommand{\defeq}{\eqqcolon}

\newcommand{\SDM}{\ensuremath{\boldsymbol{\bar{S}} \ }}
\newcommand{\SDMO}[1]{\ensuremath{\boldsymbol{\bar{S}}[{#1}]}}

\renewcommand{\sigma}{\Sigma}

\newcommand{\ii}{\mathrm{i}}

\newcommand{\T}{\mathsf{T}}
\renewcommand{\H}{\dagger}

\newcommand{\SpH}{\mathrm{Sp}^\dagger}
\newcommand{\spH}{\mathrm{sp}^\dagger}

\newcommand{\LigoMIT}{LIGO Laboratory, Massachusetts Institute of Technology,
	Cambridge, MA 02139}
\newcommand{\MechMIT}{Department of Mechanical Engineering,
	Massachusetts Institute of Technology, Cambridge, MA 02139}

\begin{document}
	
	\title{Quantum Linear Time-Translation-Invariant Systems: \\
		Conjugate Symplectic Structure, Uncertainty Bounds, and Tomography}
	
	\author{Jacques Ding}
	\email{ding@apc.in2p3.fr}
	\affiliation{\LigoMIT}
	\affiliation{Laboratoire Astroparticule et Cosmologie, Universit\'e Paris
		Cit\'e, Paris, 75000, France}
	\author{Hudson A. Loughlin}
	\affiliation{\LigoMIT}
	\author{Vivishek Sudhir}
	\email{vivishek@mit.edu}
	\affiliation{\LigoMIT}
	\affiliation{\MechMIT}
	\date{\today}
	
	\begin{abstract}
		Linear time-translation-invariant (LTI) models offer simple, yet powerful, abstractions of complex classical dynamical systems. Quantum versions of such models have so far relied on assumptions of Markovianity or an internal state-space description. We develop a general quantization scheme for multimode classical LTI systems that reveals their fundamental quantum noise, is applicable to non-Markovian scenarios, and does not require knowledge of an internal description. The resulting model is that of an open quantum LTI system whose dilation to a closed system is characterized by elements of the conjugate symplectic group. Using Lie group techniques, we show that such systems can be synthesized using frequency-dependent interferometers and squeezers. We derive tighter Heisenberg uncertainty bounds, which constrain the ultimate performance of any LTI system, and obtain an invariant representation of their output noise covariance matrix that reveals the ubiquity of ``complex squeezing'' in lossy systems. This frequency-dependent quantum resource can be hidden to homodyne and heterodyne detection and can only be revealed with more general ``symplectodyne'' detection. These results establish a complete and systematic framework for the analysis, synthesis, and measurement of arbitrary quantum LTI systems.
	\end{abstract}

	\maketitle

	\section{Introduction}
	
	A basic problem in science is to produce a ``minimal'' model of a ``black
	box'' system based on a set of response measurements on it. By a ``black
	box'', we mean that the system is only accessible through a set of input and
	output ports.
	Finally, some principle of economy dictates what we mean by ``minimal''.
	The model so inferred is the outcome of the scientific method applied to
	that system at the level of abstraction represented by the accessible
	ports.
	Systems theory formalizes the
	above task in the case where the system is classical. A high-point of this
	subject is the theory of linear time-translation-invariant (LTI) classical
	systems, which offers a simple and powerful framework to analyze 
	\cite{ZadDes63,Brock70,Kail80} 
	and synthesize \cite{Cauer58,Belev68,Youla15} complex classical systems.
	
	The purpose of this work is to present a framework for the analysis and 
	synthesis of 
	quantum LTI systems based only on their input-output description.
	Unlike a classical system, where consideration of noise is discretionary,
	any description of a quantum system has to contend with the unavoidable
	quantum noise in its accessible and inaccessible inputs. Quantum
	noises \cite{HudPar84,GardCol85,Parth92,GardZoll99} furnish the
	appropriate mathematical model for the ubiquitous and fundamental noises
	afflicting all quantum systems. In essence, quantum systems are modeled as
	being driven by quantum noises. Thus, our task is to define and
	characterize the set of linear time-translation-invariant transformations
	on quantum noise, without invoking any further assumptions. In
	particular, we do not presume knowledge of an internal description of the
	system, such as via a Hamiltonian, or quantum stochastic differential
	equations
	\cite{GardCol85,Gard93,GouYan08,JamPet08,GouJam09,Clerk10,Shaiju12,Yam14,Yam14b,CombSar17,ZhaPet18,NurYam17,Gouzien20,
		BentMiao21,BentMiao23}.
	Jettisoning that assumption makes our formalism more general --- for
	example, it can be applied to systems that do not have a state-space
	description or are non-Markovian --- and amenable to studying systems only
	based on what can be operationally verified about them. Effectively, we 
	provide
	a scheme to quantize a given input-output model of a classical LTI system, 
	and then to 
	synthesize the resulting quantum description using commonly available 
	laboratory elements.
	
	Our formalism is pertinent in the study of complex systems
	that operate at the quantum-noise limit. For example, an element as simple 
	and ubiquitous as an
	optical cavity involves a physical time delay, so that a Markovian 
	state-space
	model \cite{GardCol85} of it can be insufficient;
	such a lack is apparent in today's most sensitive and complex quantum 
	system \cite{McCuller21}.
	
	Finally, our formalism is a generalization of techniques developed in
	Gaussian quantum information 
	\cite{Arvind_1995,Braunstein05,WeedPiran12,Fabre_2020,serafini23} in the 
	sense that 
	we study transformations of time-dependent Gaussian quantum stochastic 
	processes.
	This is equivalent, as we will show, to the study of the group of
	frequency-dependent ``conjugate symplectic'' transformations which subsumes 
	the oft-studied real symplectic group
	in Gaussian quantum information. To our knowledge, this is the first 
	systematic study along these lines.
	
	The rest of this paper is structured as follows. In \cref{sec:LTIquantum}, 
	we show how an input-output LTI 
	model of a classical system can be promoted to a valid model of a quantum 
	LTI system by adding appropriate 
	noise modes. 
	The resulting open quantum LTI system can be seen as a subsystems of a 
	dilated closed quantum LTI system. 
	In \cref{sec:ConjSymp}, we show that closed quantum LTI systems are 
	elements of the conjugate symplectic Lie group. 
	We then use Lie group techniques to construct a one-parameter unitary 
	representation of this group 
	and show that arbitrary quantum LTI dynamics can be synthesized in terms of 
	optical elements readily constructed 
	in the lab. This parallels the solution of the classical synthesis problem 
	\cite{Cauer58,Belev68,Youla15} where a given
	classical LTI model is realized in terms of commonly available electrical 
	elements.
	\Cref{sec:uncertainty} establishes tight uncertainty bounds for open and 
	closed quantum LTI systems that are stricter than the usual Heisenberg 
	uncertainty relation, and elucidates invariant properties of quantum noise 
	in LTI systems.
	The latter generalizes the real-symplectic Williamson theorem 
	\cite{Williamson36} and identifies
	that quantum LTI systems generally exhibit ``complex squeezing'', a quantum 
	resource that is undetectable with homodyne or heterodyne detection, but 
	can be observed with a more general ``symplectodyne'' 
	detection [\cref{sec:realization_SDM}]. Finally, in \cref{sec:applications} 
	we explore the implications of the theory of quantum LTI systems developed 
	in the preceding sections to the Schawlow-Townes limit 
	in laser theory \cite{Loughlin23} and to quantum noise in gravitational 
	wave detectors such as LIGO.
	
	\begin{figure*}[t!]
		\centering
		\includegraphics[width=\textwidth]{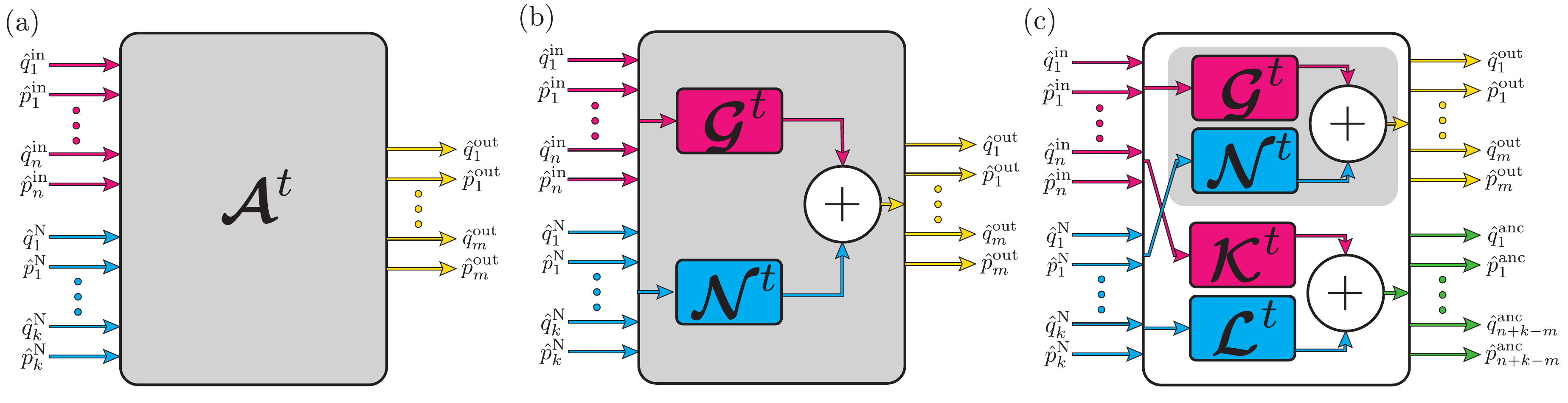}{}
		\caption{\textit{Quantum LTI System.} A quantum LTI system is described
			by $n$ input quantum noises, which are mapped to $m$ output quantum
			noises. 
			(a) In general, a quantum LTI system will map $n$ accessible inputs 
			and $k$ inaccessible inputs
			to its $m$ outputs through a map $\mathbfcal{A}$. For a linear 
			system, the maps from accessible and inaccessible inputs
			to outputs, $\mathbfcal{G}$ and $\mathbfcal{N}$, can be considered 
			separately as in (b) (see \cref{eq:xoutMN}).
			(c) By adding $n+k-m$ ancillary output modes (see 
			\cref{subsec:quantumNoiseInLtiSystems}) 
			and the maps $\mathbfcal{K}$ and $\mathbfcal{L}$
			from accessible and inaccessible inputs to ancillary outputs, 
			the map $\mathbfcal{A}$ can be dilated to a closed system (see 
			\cref{subsec:LTIdilated}).}
		\label{fig:quantumLti}
	\end{figure*}
	
	
	\section{LTI transformations of quantum noise}\label{sec:LTIquantum}
	
	\subsection{Quantum LTI systems}
	
	We begin by studying ``black box'' systems with a known relationship 
	between their inputs
	and outputs but without assuming any knowledge of their internal dynamics.
	We will show that by making the minimal assumptions of linearity and 
	time-shift invariance, 
	these systems are fully characterized by a set of ``transfer functions'' 
	that dictate both their
	response to classical fields and to quantum noise.
	
	%
	To wit, consider a classical system that maps some set of $n$ inputs, 
	$\boldsymbol{x}^\text{in}(t) =
	\{x_i^\t{in}(t)\}_{i=1,\ldots,n}$ to some set of $m$ outputs 
	$\boldsymbol{x}^\text{out}(t) = \{x_i^\t{out}(t)\}_{i=1,\ldots,m}$
	through a dynamical map $\mathbfcal{G}^t$ at each time $t$ via
	\begin{equation}\label{eq:classicalInOut}
		\boldsymbol{x}^\text{out}(t) = 
		\mathbfcal{G}^t[\boldsymbol{x}^\text{in}].
	\end{equation}
	As far as the input-output description is concerned, the system \emph{is} 
	the
	set of maps $\{\mathbfcal{G}^t\}$.
	
	We now make two key assumptions: that the dynamical maps are \emph{linear}
	and \emph{time-translation invariant}. These notions are defined precisely 
	as
	follows.
	\begin{definition}\label{def:linear_classical}
		A classical system $\{\mathbfcal{G}^t\}$ is linear if
		\begin{equation}\label{eq:classicalLinearity}
			\mathbfcal{G}^t[\lambda \boldsymbol{x} + \lambda' 
			\boldsymbol{x}^\prime] = \lambda
			\mathbfcal{G}^t[\boldsymbol{x}] + \lambda' 
			\mathbfcal{G}^t[\boldsymbol{x}^\prime].
		\end{equation}
		for all square-integrable functions 
		$\boldsymbol{x},\boldsymbol{x'}$
		and all constants $\lambda,\lambda'$ at each instant $t$.
	\end{definition}
	In general, a linear system's map, $\mathbfcal{G}^t$ from its $n$ inputs to 
	its $m$ outputs can be decomposed in terms of a collection of maps, 
	$\mathbfcal{G}^t_{ij}$ from its $j^\text{th}$ input port to its 
	$i^\text{th}$ output port (see \cref{thm:MimoToSiso}). 
	\begin{definition}\label{def:LTI}
		Given the time-translation operator $\mathcal{T}^s$, defined by, 
		$\mathcal{T}^s[f(t)] \eqdef f(t-s)$,
		the classical linear system $\{\mathbfcal{G}^t\}$ 
		is said to be linear time-translation-invariant (LTI) if its dynamical 
		map
		commutes with the time-translation operator, i.e.
		\begin{equation}\label{eq:classicalTimeShiftInvar}
			(\mathcal{T}^s \circ \mathcal{G}^t_{ij})[x] = (\mathcal{G}^t_{ij}
			\circ \mathcal{T}^s)[x],
		\end{equation}
		for all inputs $x$, input and output ports, $i,j$, and times $t,s$.
	\end{definition}
	
	The assumption of linearity is well-justified for a large class of systems 
	since any system will
	respond approximately linearly to sufficiently small signals (as long as its
	dynamical map's first derivatives with respect to these signals are 
	non-vanishing, which defines
	the linear operating point).
	Systems that do not have an internal time reference, i.e. are autonomous in 
	their internal mechanism,
	satisfy the assumption of time-translation invariance.
	Together, the LTI assumptions imply several key properties, of
	which the most pertinent is that the input-output response in 
	\cref{eq:classicalInOut} can be
	put in the form
	\begin{equation}\label{eq:classicalInOutM}
		\boldsymbol{x}^\text{out}[\omega] = \boldsymbol{G}[\omega] 
		\boldsymbol{x}^\text{in}[\omega],
	\end{equation}
	in terms of the Fourier transforms of the inputs and outputs (see 
	\cref{app:ltiClassical} for a proof), and the transfer function matrix 
	$\boldsymbol{G}[\omega]$ is
	uniquely related to $\{\mathbfcal{G}^t\}$. 
	
	To extend the notion of a LTI system from the classical to quantum domain,
	we note that inputs and outputs to a quantum system, in contrast to 
	classical ones,
	need to be described by operator-valued stochastic processes. A convenient 
	mathematical
	model is that of quantum noises defined as follows, and realized in 
	practice by the
	quadratures of traveling electromagnetic fields \cite{Blow_Loudon}, or 
	voltages and currents
	in a transmission line \cite{Yurke84,Clerk10,Blais21}, typically used to 
	drive and probe
	quantum optical or electronic systems.
	
	\begin{definition}
		Quantum noises are pairs of operator-valued stochastic processes
		$\{\hat{q}_i(t),\hat{p}_i(t)\}_{i=1,\ldots,n}$ that satisfy
		\begin{equation}\label{eq:qwnCommutators}
			\begin{split}
				[\hat{q}_i(t), \hat{q}_j(t')] &= 0 \\
				[\hat{p}_i(t), \hat{p}_j(t')] &= 0 \\
				[\hat{q}_i(t), \hat{p}_j(t')] &= \ii \delta_{ij} \delta(t-t').
			\end{split}
		\end{equation}
	\end{definition}
	
	By a quantum LTI system we have in mind the scenario depicted in 
	\cref{fig:quantumLti}:
	a set of $n$ input ports and $m$ output ports are accessible for classical 
	response measurements ---
	i.e. the inputs are driven by classical signals and their classical output 
	response is measured;
	the observed classical response of the system satisfies the LTI assumptions
	in \cref{eq:classicalLinearity,eq:classicalTimeShiftInvar}, i.e.
	\begin{equation}
		\boldsymbol{x}^\text{out}(t) = 
		\mathbfcal{G}^t[\boldsymbol{x}^\text{in}],
	\end{equation}
	where $\boldsymbol{x}\eqdef \avg*{\boldsymbol{\hat x}}$ denotes the 
	expectation values of the input
	$\boldsymbol{\hat{x}}^\text{in}(t) \eqdef
	[\hat{q}^\text{in}_1(t),...,\hat{q}^\text{in}_n(t),\hat{p}^\text{in}_1(t),...,\hat{p}^\text{in}_n(t)]^T$
	and output $\boldsymbol{\hat{x}}^\text{out}(t) \eqdef
	[\hat{q}^\text{out}_1(t),...,\hat{q}^\text{out}_m(t),\hat{p}^\text{out}_1(t),...,\hat{p}^\text{out}_m(t)]^T$.
	Since this definition only relies on the expectation values of quantum
	noises, we can determine whether a quantum system is LTI through
	classical measurements.
	Once a quantum system is verified to be LTI in this sense, the implications 
	of classical systems theory apply to the above input-output equation. 
	It can be put in the form of \cref{eq:classicalInOutM} 
	which governs how the quantum LTI system maps its inputs to outputs 
	\textit{in 
		expectation value only}.
	As a result, we can use a classical measurement to
	determine the transfer matrix $\boldsymbol{G}[\omega]$.
	
	The analysis so far omits quantum fluctuations in the accessible inputs and
	in any inaccessible inputs.
	That is, \cref{eq:classicalInOutM} cannot tacitly be promoted to an 
	operator 
	relation
	(for example if $\{\mathbfcal{G}^t\}$ does not arise from unitary dynamics).
	The output ($\boldsymbol{x}^\t{out}$) can still be modeled as arising from 
	the
	accessible ($\boldsymbol{x}^\t{in}$) and some additional inaccessible 
	($\boldsymbol{x}^\t{N}$) inputs, i.e.
	$\boldsymbol{x}^\text{out} = 
	\mathbfcal{A}[\boldsymbol{x}^\text{in},\boldsymbol{x}^\text{N}]$,
	where $\mathbfcal{A}$ maps the accessible and inaccessible inputs to the 
	accessible outputs.
	Assuming this map is linear, then
	\begin{equation}\label{eq:xoutMN}
		\begin{split}
			\boldsymbol{x}^\text{out} 
			= \mathbfcal{A}[\boldsymbol{x}^\text{in},0] + 
			\mathbfcal{A}[0,\boldsymbol{x}^\text{N}] 
			\defeq \mathbfcal{G}[\boldsymbol{x}^\text{in}] + 
			\mathbfcal{N}[\boldsymbol{x}^\text{N}],
		\end{split}
	\end{equation}
	where $\mathbfcal{N}$ maps the inaccessible inputs to the accessible 
	outputs.
	Further, if $\mathbfcal{A}$ is assumed time-translation invariant, so is 
	$\mathbfcal{N}$,
	and it admits a frequency domain representation just as $\mathbfcal{G}$ 
	does.
	Thus, \cref{eq:classicalInOutM} can be promoted to operator relations at 
	the expense of adding 
	additional quantum noises:
	\begin{equation}\label{eq:quantumLtiOp}
		\boldsymbol{\hat{x}}^\text{out}[\omega] = \boldsymbol{G}[\omega]
		\boldsymbol{\hat{x}}^\text{in}[\omega] + \boldsymbol{N}[\omega]
		\boldsymbol{\hat{x}}^\text{N}[\omega]
	\end{equation}
	where $\boldsymbol{\hat{x}}^\text{N}(t) \eqdef
	[\hat{q}^\text{N}_1(t),...,\hat{q}^\text{N}_k(t),\hat{p}^\text{N}_1(t),...,\hat{p}^\text{N}_k(t)]^T$
	are $k$ additional quantum noises with zero mean so that in an expectation 
	value sense, \cref{eq:quantumLtiOp} reduces
	to \cref{eq:classicalInOutM}.
	We call $\boldsymbol{x}^\t{N}$ ``noise modes'' to emphasize that their 
	primary effect is to add noise in
	excess of what is conveyed by the accessible inputs alone. 
	\Cref{fig:quantumLti}(a,b) shows the relation between
	the linear map $\mathbfcal{A}$ from all inputs to the accessible outputs 
	and the maps $\mathbfcal{G}$ from accessible
	inputs to accessible outputs and $\mathbfcal{N}$ from inaccessible inputs 
	to accessible outputs.
	
	\subsection{Quantum noise in LTI 
	systems}\label{subsec:quantumNoiseInLtiSystems}
	
	Classical response measurements only reveal $\boldsymbol{G}$; 
	$\boldsymbol{N}$ remains hidden
	since the ports corresponding to those inputs are inaccessible, i.e.
	the noise modes $\boldsymbol{\hat{x}}^\t{N}$ are internal to the system.
	
	As we now show, even without an internal model of the system, it is possible
	to determine a class of hidden matrices $\boldsymbol{N}$ that ensure the 
	consistent
	mathematical description of the quantum LTI system. Further, as we will 
	show, this
	class determines the \emph{minimum} quantum noise that must be added given 
	the
	accessible description of a quantum LTI system characterized by 
	$\boldsymbol{G}$.
	
	We require that the system's output $\boldsymbol{\hat{x}}^\t{out}(t)$ obey 
	the
	canonical commutation relations (CCR) in \cref{eq:qwnCommutators}; in the 
	frequency domain
	they read
	\begin{equation}\label{eq:ccrAndJdef}
		\begin{split}
			\left[ \hat{x}_i[\omega], \hat{x}_j[\omega^\prime]^\dagger \right]
			&= 2 \pi \ii \, \delta[\omega - \omega^\prime] J_{ij}^\t{out},\\
			\t{where}\qquad \boldsymbol{J}^\t{out} &\eqdef
			\begin{bmatrix} \boldsymbol{0}_m & \boldsymbol{1}_m \\ 
			-\boldsymbol{1}_m &
				\boldsymbol{0}_m\end{bmatrix}
		\end{split}
	\end{equation}
	is the $2m$-dimensional symplectic matrix of the $m$ output modes, given in
	in terms of $\boldsymbol{1}_m$ and $\boldsymbol{0}_m$, the $m$-dimensional 
	identity and zero
	matrices respectively.
	Stipulating that the outputs are given by the system's input-output relation
	[\cref{eq:quantumLtiOp}] then implies that
	\begin{equation}\label{eq:ccrMatrixConstraint}
		\boldsymbol{N}[\omega] \ii \boldsymbol{J}^\text{N} 
		\boldsymbol{N}[\omega]^\dagger =
		\ii
		\boldsymbol{J}^\text{out} - \boldsymbol{G}[\omega] \ii 
		\boldsymbol{J}^\text{in}
		\boldsymbol{G}[\omega]^\dagger,
	\end{equation}
	where $\boldsymbol{J}^\text{in}, \boldsymbol{J}^\text{N}$ are respectively 
	the $2n$ and
	$2k$ dimensional symplectic matrices corresponding to the inputs and noise 
	modes respectively.
	\Cref{eq:ccrMatrixConstraint} is sufficient to determine the mathematical 
	consistency of any hidden transfer
	matrix $\boldsymbol{N}$.

	Since the left and right hand sides of \cref{eq:ccrMatrixConstraint} are 
	both
	Hermitian, they can be unitarily diagonalized at every frequency as
	\begin{equation}
		\ii \boldsymbol{J}^\text{out} - \boldsymbol{G}[\omega] \ii 
		\boldsymbol{J}^\text{in}
		\boldsymbol{G}[\omega]^\dagger = \boldsymbol{U}[\omega] \begin{bmatrix}
			\boldsymbol{D}^+[\omega] & \boldsymbol{0} & \boldsymbol{0} \\ 
			\boldsymbol{0}
			&\boldsymbol{D}^-[\omega] & \boldsymbol{0} \\ \boldsymbol{0} & 
			\boldsymbol{0} &
			\boldsymbol{0} \end{bmatrix} \boldsymbol{U}[\omega]^\dagger,
	\end{equation}
	where $\boldsymbol{D}^+[\omega]$ and
	$\boldsymbol{D}^-[\omega]$ are diagonal matrices with only positive and 
	negative
	entries respectively. In general, $\boldsymbol{D}^+$ and $\boldsymbol{D}^-$ 
	can have different dimensions;
	we denote $d_\pm \eqdef \dim(\boldsymbol{D}^\pm[\omega])$. This dimension 
	determines the minimum number of
	noise modes as per the following theorem (see SI 
	Section I for the proof).
	
	\begin{theorem}\label{thm:min_noise_modes}
		The minimum number of noise modes for a quantum LTI system that 
		satisfies
		\Cref{eq:ccrMatrixConstraint} is $\ell = \t{max}(d_-, d_+)$.
	\end{theorem}
	
	While \cref{eq:ccrMatrixConstraint} cannot be satisfied with fewer than 
	$\ell$
	noise modes, it can always be satisfied by adding more that $\ell$ noise modes, 
	or example by taking the minimal-noise construction and adding amplifiers and 
		attenuators in series, or by decomposing a single loss as a
		series of losses.
	Thus, the bound in \cref{thm:min_noise_modes} is tight.
	
	In the case that we add exactly $\ell$ noise modes, the matrix
	$\boldsymbol{N}[\omega]$ is given by
	\begin{equation}\label{eq:optNfreqDep}
		\boldsymbol{N}[\omega] = \boldsymbol{U}[\omega] 
		\boldsymbol{\Gamma}[\omega]
		\boldsymbol{P}^\dagger,
	\end{equation}
	where $\boldsymbol{\Gamma}[\omega] = \{\gamma_{i}[\omega] \delta_{ij}\}$
	is a $m \times k$ matrix with diagonal entries
	\begin{equation}\label{eq:singularValuesN}
		\gamma_{i} = \begin{cases}
			\sqrt{D^+_{ii}},& \text{if } 1 \leq i \leq d_+\\
			\sqrt{|D^-_{ii}|},& \text{if } d_+ +1 \leq i \leq d_+ + d_-\\
			0, & \text{otherwise}
		\end{cases}
	\end{equation}
	and all other entries zero, and
	$\boldsymbol{P}$ is the unitary matrix that diagonalizes 
	$i\boldsymbol{J}^\text{N}$.
	Note that $\gamma_i$ are the singular values of $\boldsymbol{N}$.
	These should be ordered such that the singular
	values determined by $\boldsymbol{D}^+$ multiply $+1$ in the 
	diagonalization
	of $\ii \boldsymbol{J}^\text{N}$ and the singular values determined by 
	$\boldsymbol{D}^-$
	multiply $-1$.
	
	If $\boldsymbol{G}[\omega]$ and $\boldsymbol{N}[\omega]$ are
	frequency-independent \cite{Braunstein05,WeedPiran12}, their entries must 
	be real-valued
	(this follows from the fact that each component of $\boldsymbol{x}(t)$ is 
	Hermitian,
	i.e.  $\hat{x}_i[\omega]^\dagger = \hat{x}_i[-\omega]$).
	In this case, both the left and
	right-hand sides of \cref{eq:ccrMatrixConstraint} are Hermitian and
	anti-symmetric, so their eigenvalues must be real-valued and come in
	positive and negative pairs, implying $d_+ = d_-$, and so
	the minimum number of added noise modes is $\ell = d_\pm$.
	That is, in the frequency-independent case, at most the same number of
	noise modes must be added as there are outputs.
	By contrast, in the frequency-dependent case, up to
	twice as many noises modes as outputs need to be added.
	
	We will return to the important question of how to quantify the added noise 
	at 
	the accessible outputs of the system,
	the minimum added noise, and any trade-offs therein, in 
	\cref{sec:uncertainty}.
	
	\subsection{Dilation of a quantum LTI system}\label{subsec:LTIdilated}
	
	The quantum system resulting from the addition of noise modes is still not 
	unitary.
	A necessary condition to realize a unitary system is that the total number 
	of
	inputs and outputs be equal. The quantum LTI system in 
	\cref{eq:quantumLtiOp} can be
	enlarged, or dilated, into the form (shown in \cref{fig:quantumLti}c)
	\begin{equation}\label{eq:extendedLti}
		\begin{bmatrix} \boldsymbol{\hat{x}}^\text{out}[\omega] \\
			\boldsymbol{\hat{x}}^\text{anc}[\omega] \end{bmatrix} = 
			\begin{bmatrix}
			\boldsymbol{G}[\omega] & \boldsymbol{N}[\omega] \\ 
			\boldsymbol{K}[\omega] &
			\boldsymbol{L}[\omega] \end{bmatrix} \begin{bmatrix}
			\boldsymbol{\hat{x}}^\text{in}[\omega] \\ 
			\boldsymbol{\hat{x}}^\text{N}[\omega]
		\end{bmatrix},
	\end{equation}
	which includes ancillary outputs ($\boldsymbol{\hat{x}}^\text{anc}$)
	so as to ensure that the total number of outputs equals the total number of 
	inputs.
	Here, $\boldsymbol{\hat{x}}^\text{in}, \boldsymbol{\hat{x}}^\text{out}$ are 
	the system's accessible
	inputs and outputs, while $\boldsymbol{\hat{x}}^\text{N}, 
	\boldsymbol{\hat{x}}^\text{anc}$ are the system's
	inaccessible, or unobserved, inputs and outputs.
	The matrices $\boldsymbol{K}$ and $\boldsymbol{L}$ represent the LTI maps 
	$\mathbfcal{K}$ and $\mathbfcal{L}$ that respectively map the system's 
	accessible and inaccessible inputs to its inaccessible outputs,
	as depicted in \cref{fig:quantumLti}.
	
	Importantly, by construction, the matrix
	\begin{equation}\label{eq:Mext}
		\boldsymbol{M}^\text{ext}[\omega] \eqdef \begin{bmatrix}
			\boldsymbol{G}[\omega] & \boldsymbol{N}[\omega] \\ 
			\boldsymbol{K}[\omega] &
			\boldsymbol{L}[\omega] \end{bmatrix}
	\end{equation}
	is square, and the dilated input-output relation in \cref{eq:extendedLti} 
	can be cast into the form
	\begin{equation}\label{eq:ltiEnlarged}
		\boldsymbol{\hat{y}}^\text{out}[\omega] = \boldsymbol{M}^\t{ext}[\omega]
		\boldsymbol{\hat{y}}^\text{in}[\omega],
	\end{equation}
	of which, the system given in \cref{eq:quantumLtiOp} (and shown in 
	\cref{fig:quantumLti}b) is an open
	subsystem. As before, the inputs and outputs of this dilated system must 
	satisfy the CCRs, which implies
	that
	\begin{equation}\label{eq:MextSymp}
		\boldsymbol{M}^\text{ext}[\omega] \boldsymbol{J}^\text{ext}
		\boldsymbol{M}^\text{ext}[\omega]^\dagger = \boldsymbol{J}^\text{ext}
	\end{equation}
	where
	\begin{equation}
		\boldsymbol{J}^\text{ext} \eqdef \begin{bmatrix}
			\boldsymbol{J}_{2n \times 2n} & \boldsymbol{0} \\ \boldsymbol{0} & 
			\boldsymbol{J}_{2k
				\times 2k} \end{bmatrix}.
	\end{equation}
	Any square matrix $\boldsymbol{M}^\t{ext}$ that satisfies 
	\cref{eq:MextSymp} forms a group --- the
	conjugate symplectic group.
	This group plays a key role in the study of frequency-dependent 
	transformations of quantum noise,
	and subsumes the symplectic group that arises in the study of
	Gaussian quantum random variables.
	The following section explicates the relevant properties of the conjugate
	symplectic group. In particular, \cref{eq:MextSymp} is sufficient to ensure 
	that the dilated input-output
	relation in \cref{eq:ltiEnlarged} arises from reversible and therefore 
	unitary, but not necessarily Markovian,
	quantum dynamics.
	
	The outstanding question in the construction of the dilated system 
	described by $\boldsymbol{M}^\t{ext}$
	[\cref{eq:Mext}] is whether, given the matrices 
	$\boldsymbol{G},\boldsymbol{N}$
	(the latter possibly constructed from the former
	via the procedure in \cref{subsec:quantumNoiseInLtiSystems}), it is always 
	possible to
	find $\boldsymbol{K},\boldsymbol{L}$
	that enable a unitary dilation. A conjugate-symplectic diagonalization 
	procedure, extending a result in the frequency-independent case 
	\cite{Silva08}, shows that it is always possible to find 
	$\boldsymbol{K},\boldsymbol{L}$
	for a unitary dilation (see SI Section II) 
	\footnote{
		In Supplementary Information Section II, we 
		prove that a quantum LTI system with $n$ input modes, $m$ output modes 
		and $k$ noise modes can be considered to be a subsystem of a dilated 
		unitary system with $n+k-m$ ancillary output modes. The dilated system 
		will have the same mean dynamics of the map from the $n$ observed inputs to 
		the $m$ observed outputs. Ref. \cite{Silva08} proves a stronger dilation 
		theorem for real-symplectic systems: a Gaussian bosonic channel from $n$ 
		inputs to $n$ outputs can be dilated to a unitary map by adding $2n$ 
		additional modes, such that both the system's mean response and its output 
		noise is preserved in the dilated system.
		We leave the proof of the full Gaussian bosonic channel dilation theorem 
		to future work. However, we conjecture that a LTI Gaussian bosonic 
		channel from $n$ inputs to $n$ outputs can be dilated to a unitary map 
		by adding $4n$ additional modes. We anticipate that the additional 
		factor of $2$ compared to the frequency-independent case comes from 
		lifting the anti-symmetry condition, as in 
		\cref{subsec:quantumNoiseInLtiSystems}.
	}.
	In effect, any LTI system can be thought of as a subsystem of
	a dilated unitary system governed by \cref{eq:ltiEnlarged} with some of the 
	output modes
	($\boldsymbol{\hat{x}}^\t{anc}$) inaccessible.
	
	
	\section{Conjugate symplectic group}\label{sec:ConjSymp}
	
	Following the discussion so far, it is sufficient to study a quantum
	LTI ``black box'' with $n$ input quantum noises 
	$\boldsymbol{\hat{x}}^\text{in}(t)$ and $n$
	outputs $\boldsymbol{\hat{x}}^\text{out}(t)$, with the complex-valued 
	transfer
	function matrix $\boldsymbol{M}[\omega]$ that describes the input-output 
	relation
	\begin{equation}\label{eq:transfer_function_matrix_def}
		\boldsymbol{\hat{x}}^\text{in}[\omega] \mapsto
		\boldsymbol{\hat{x}}^\text{out}[\omega] = \boldsymbol{M}[\omega]
		\boldsymbol{\hat{x}}^\text{in}[\omega].
	\end{equation}
	Recall that we assume only the following minimal hypotheses:
	(i) the quantum noises are hermitian, i.e. $\hat{x}_i^\dagger(t) = 
	\hat{x}_i(t)$, for all $i\in \{1,\ldots, 2n\}$;
	and, (ii) they satisfy the canonical commutation relations given in 
	\cref{eq:qwnCommutators}.
	The first condition implies that
	\begin{equation}\label{eq:freq_domain_condition}
		\boldsymbol{M}[\omega]^*  = \boldsymbol{M}[-\omega];
	\end{equation}
	in particular $\boldsymbol{M}[0] \in \mathcal{M}_{2n}(\mathbb{R})$, i.e. it 
	is a $2n$-dimensional
	real matrix.
	The second condition implies that
	\begin{equation} \label{eq:conjugate_symp_condition}
		\boldsymbol{M}[\omega]\;  \boldsymbol{J \; M}[\omega]^\dagger =
		\boldsymbol{J}
	\end{equation}
	where $\boldsymbol{J}$ is the $2n$-dimensional symplectic matrix.
	These physical requirements motivate the following mathematical definition 
	of the
	conjugate symplectic group \footnote{The
		conjugate symplectic group $\SpH(2n)$ is not to be confused with the 
		complex symplectic group
		$\text{Sp}(2n,\mathbb{C}) = \left\{ \boldsymbol{M}\in
		\mathcal{M}_{2n}(\mathbb{C})\mid
		\boldsymbol{M}\boldsymbol{J}\boldsymbol{M}^T = \boldsymbol{J}\right\}$;
		see remark 2.7 of \cite{edelman2022fifty} and references therein.}.
	
	\begin{definition}
		The conjugate symplectic group of order $2n$ is the set
		\begin{equation}
			\SpH(2n) \eqdef \left\{ \boldsymbol{M} \in
			\mathcal{M}_{2n}(\mathbb{C}) \mid \boldsymbol{M J M}^\H =
			\boldsymbol{J} \right\}.
		\end{equation}
		\noindent where $\boldsymbol{J}$ is the $2n$-dimensional symplectic 
		matrix.
	\end{definition}
	
	That the set $\SpH(2n)$ is a group can be directly verified (using the fact 
	that
	$\boldsymbol{J}^\H = \boldsymbol{J}^{-1} = -\boldsymbol{J}$). In fact, the 
	transfer matrix $\boldsymbol{M}[\omega]$ at each frequency
	is an element of the conjugate symplectic group.
	It can be verified that the condition in \cref{eq:freq_domain_condition} is 
	compatible with the group structure, i.e. if
	$\boldsymbol{M}[\omega]$ is conjugate symplectic, then so is 
	$\boldsymbol{M}[-\omega]^*$.
	Thus, \cref{eq:freq_domain_condition} may be viewed as defining the 
	negative-frequency extension of the conjugate
	symplectic, positive-frequency transfer matrix.
	
	The group structure implies in particular that if $\boldsymbol{M}[\omega]$
	is conjugate symplectic, so is $\boldsymbol{M}^{-1}[\omega]$; i.e. the 
	input-output relation in
	\cref{eq:transfer_function_matrix_def} is reversible \footnote{We 
	distinguish a 
		transformation $\boldsymbol{M}[\omega]$ which is reversible, i.e. for 
		which a 
		matrix inverse $\boldsymbol{M}^{-1}[\omega]$ exists from a mathematical 
		standpoint only, from a transformation which is \textit{causally} 
		reversible, 
		i.e. for which $\boldsymbol{M}^{-1}[\omega]$ exists and respects 
		causality.}.
	This reversibility suggests that any transfer function matrix can be 
	implemented unitarily. 
	We construct the unitary implementation in \cref{sec:LieUnitaryRep} and its
	synthesis --- using common laboratory optical elements --- in 
	\cref{sec:OptDecomp}.
	
	\subsection{One-parameter unitary representation}\label{sec:LieUnitaryRep}
	
	In order to construct a unitary representation of $\SpH(2n)$ in the Hilbert 
	space $\mathscr{H}$ of the
	quantum noises, we pass through the Lie algebra $\spH(2n)$. In what 
	follows, 
	we suppress the frequency argument unless required.
	
	In general, for matrix Lie groups with coefficients in $\mathbb{R}$ or 
	$\mathbb{C}$, the algebra is the tangent space at the 
	identity element $\boldsymbol{1}$ of the group. 
	Let $\tau \mapsto \boldsymbol{M}_\tau$ be a continuous map from the reals 
	$\{\tau\}=\mathbb{R}$ to the group such that 
	$\boldsymbol{M}_0 = \boldsymbol{1}$. The group property implies that 
	$\boldsymbol{M}_\tau = (\boldsymbol{M}_{\tau/m})^m$ for any integer $m$.
	In the neighborhood of the identity, $\boldsymbol{M}_{\tau} \overset{\tau 
	\to 0}{=} 
	\boldsymbol{1} + \tau\boldsymbol{\Lambda}$, where $\boldsymbol{\Lambda} \in 
	\mathcal{M}_{2n}(\mathbb{C})$ is the generator of the algebra 
	\footnote{Note that here we follow the 
		mathematician's convention for defining a generator, which defers from 
		the physicist's convention by an $\ii$.}.
	
	The group element can then be expressed in terms of the generator:
	\begin{equation*}
		\begin{split}
			\boldsymbol{M}_\tau = \lim_{m \rightarrow \infty} 
			(\boldsymbol{M}_{\tau/m})^m
			\approx \lim_{m \rightarrow \infty} \left(\boldsymbol{1} + 
			\frac{\tau}{m}\boldsymbol{\Lambda}\right)^m
			= e^{\tau \boldsymbol{\Lambda}}.
		\end{split}
	\end{equation*}
	Since the generator needs to generate the group in question, 
	$\boldsymbol{M}_\tau \boldsymbol{J}
	\boldsymbol{M}_\tau^\dagger = \boldsymbol{J}$; taking the derivative of 
	this 
	relation at $\tau = 0$
	gives $\boldsymbol{\Lambda J} = (\boldsymbol{\Lambda J})^\dagger$,
	which defines the conjugate symplectic algebra \footnote{
		An equivalent characterization is obtained by the (bijective) 
		transformation
		$\boldsymbol{\Lambda} \rightarrow \boldsymbol{J \Lambda}$, in which 
		case,
		$\spH(2n) \eqdef \left\{ \boldsymbol{J \Lambda} \in
		\mathcal{M}_{2n}(\mathbb{C}) \mid \boldsymbol{\Lambda} \textnormal{ is 
		hermitian} 
		\right\}$, and $e^{\tau \boldsymbol{J \Lambda}} \in \SpH(2n)$.
	}.
	
	\begin{definition}
		The conjugate symplectic Lie algebra of order $2n$ is the set
		\begin{equation}\label{eq:spH}
			\begin{split}
				\spH(2n) & \eqdef \left\{ \boldsymbol{\Lambda} \in
				\mathcal{M}_{2n}(\mathbb{C}) \mid \boldsymbol{\Lambda J} 
				\textnormal{ 
					is hermitian} \right\}.
			\end{split}
		\end{equation}
	\end{definition}

	In order to imbue physical meaning into this description, we now look for a 
	unitary representation of $\SpH(2n)$. This means a map
	\begin{align}\label{eq:unitary_representation}
		\begin{split}
			\SpH(2n) &\to \mathscr{H}\\
			\boldsymbol{M} &\mapsto \hat{U}(\boldsymbol{M}),
		\end{split}
	\end{align}
	that is unitary, i.e. $\hat{U}(\boldsymbol{M})^\dagger 
	\hat{U}(\boldsymbol{M}) = \hat{1}$, and
	respects the group structure, i.e.
	$\hat{U}(\boldsymbol{M M'}) = \hat{U}(\boldsymbol{M}) 
	\hat{U}(\boldsymbol{M'})$.  
	Arguments similar to the one above can be deployed on the representation. 
	To wit,
	the group identity is mapped to the identity element in Hilbert space, and 
	so 
	in the neighborhood of the identity,
	$\hat{U}(\boldsymbol{M}_{\tau}) \overset{\tau \to 0}{=} 
	\hat{U}(\boldsymbol{1}+ \ii \tau \boldsymbol{\Lambda})
	\approx \hat{1} + \ii \tau \hat{H}(\boldsymbol{\Lambda})$; here 
	$\boldsymbol{\Lambda} \mapsto
	\hat{H}(\boldsymbol{\Lambda})$ is a representation of the algebra 
	$\spH(2n)$ on the Hilbert space.
	By the representation property, $\hat{U}(\boldsymbol{M}_\tau) =
	\hat{U}(\boldsymbol{M}_{\tau/m})^m$ for any $m\in\mathbb{N}$; using the 
	infinitesimal form above, and taking the limit $m \rightarrow \infty$
	then gives $\hat{U}(\boldsymbol{M}_\tau) = \exp [\ii \tau 
	\hat{H}(\boldsymbol{\Lambda})]$. For $\hat{U}$ to be
	unitary, $\hat{H}(\boldsymbol{\Lambda})$ has to be hermitian \footnote{
		The argument so far has tacitly assumed that any group element can be 
		represented in the form 
		$e^{\tau \boldsymbol{\Lambda}}$; this is true since the exponential map 
		is 
		surjective in $\SpH(2n)$ \cite{yakubovich1975linear,DJOKOVIC198076}.
		We have also not been careful about the fact that in the Hilbert space 
		of quantum states, the unitary representation is 
		up to a phase (i.e. a projective unitary representation); however, this phase is
		irrelevant in any observable. 
	}.
	
	The point of the unitary representation is that the transformation 
	$\boldsymbol{M}_\tau$ must be expressible as:
	\begin{equation}
		\hat{x}_j \mapsto \hat{x}_j' = (\boldsymbol{M}_\tau)_{jk}\hat{x}_k
		= \hat{U}(\boldsymbol{M}_\tau)^\dagger \hat{x}_j 
		\hat{U}(\boldsymbol{M}_\tau).
	\end{equation}
	Evaluating these two expressions for an infinitesimal transformation gives
	(restoring the frequency arguments)
	\begin{equation}
		-\ii\, [\hat{H}(\boldsymbol{\Lambda}), \hat{x}_j[\omega]] = 
		\Lambda_{jk}[\omega]\hat{x}_k[\omega],
	\end{equation}
	which determines $\hat{H}(\boldsymbol{\Lambda})$ for a given 
	$\boldsymbol{\Lambda}$.
	The solution is
	\begin{equation}
		\begin{split}
			\hat{H}(\boldsymbol{\Lambda}) &= \frac{1}{2}\int \hat{x}_i[\omega] 
			J_{ij}
			\Lambda_{jk}[\omega] \hat{x}_k[\omega] \,\frac{\dd \omega}{2\pi}\\
			&= \frac{1}{2}\int \hat{x}_i(t) J_{ij}\Lambda_{jk}(t-t')
			\hat{x}_k(t')\, \dd t\, \dd t'.
		\end{split}
	\end{equation}
	
	From the perspective of quantum mechanics, $\hat{U}(\boldsymbol{M})$ is the
	unitary operator that realizes the dynamics represented by the black box
	transfer matrix $\boldsymbol{M}$; then, $\hat{H}(\boldsymbol{\Lambda})$, 
	being the hermitian generator of
	the unitary, is the Hamiltonian.
	Its form can be interpreted as the input quantum noise $\hat{x}_i^\t{in}$ 
	at time $t$ interacting with
	$\hat{x}_k^\t{in}$ at time $t'$ with the coupling constant 
	$J_{ij}\Lambda_{jk}(t-t')$.
	That is, the interaction is delocalized in time, encoding the potential 
	non-Markovian character of the dynamics.
	The Markovian limit corresponds to $\Lambda_{jk}(t-t') = \Lambda_{jk} 
	\delta(t-t')$.
	
	\begin{figure*}[t!]
		\centering
		\includegraphics[width=\textwidth]{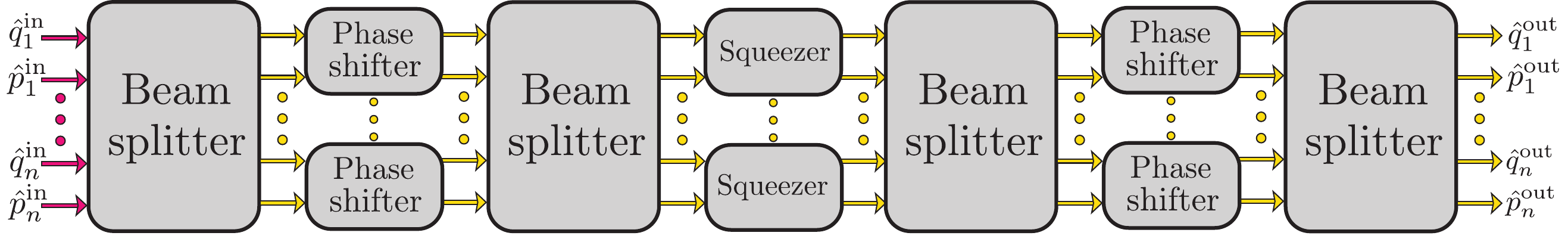}
		\caption{Optical decomposition (\cref{thm:opt_decomp}) of a 
			frequency-dependent LTI quantum system. While formally similar to 
			the 
			Bloch-Messiah decomposition on the real symplectic group 
			$\text{Sp}(2n, 
			\mathbb{R})$ \cite{Bloch62}, note that each component is a function of frequency 
			and the 
			beam-splitter component is complex-valued.}
		\label{fig:Optical_decomposition}
	\end{figure*}
	
	\subsection{Synthesis of $\SpH(2n)$ transformation}\label{sec:OptDecomp}
	
	The Hamiltonian formally conveys the microscopic structure of the
	interactions within the black box, but it does not directly inform how a 
	given
	transfer matrix $\boldsymbol{M}[\omega]$ may be synthesized in practice in 
	a lab.
	
	Any synthesis procedure is the result of factorizing a given 
	$\boldsymbol{M}\in \SpH(2n)$ in terms of other 
	elements of the conjugate symplectic group, so that each factor can be 
	physically realized.
	We prove that such a factorization can be achieved by first using a 
	conjugate symplectic singular value 
	decomposition \cite{XU20031,edelman2022fifty}, which decomposes 
	$\boldsymbol{M}$ into a product
	of elements of $\SpH(2n)\cap U(2n)$ and a positive repeating diagonal 
	matrix, followed by a 
	``cosine-sine'' decomposition in $\SpH(2n)\cap U(2n)$.
	The resulting factorization, which we call the ``optical decomposition'', 
	is as follows.
	(See Supplementary Information Section III 
	for the proof, and a sketch of its relation to the ``KAK'' decomposition in 
	the theory
	of Lie groups \cite{Knapp02}.)

	\begin{widetext}
		\begin{theorem}[Optical decomposition of 
		$\SpH(2n)$]\label{thm:opt_decomp}
			Any $\boldsymbol{M}\in \SpH(2n)$ can be uniquely factorized as
			\begin{equation}\label{eq:optical_decomposition}
				\boldsymbol{M} = 
				\begin{bmatrix}
					\boldsymbol{V}_1 & \boldsymbol{0} \\
					\boldsymbol{0} & \boldsymbol{V}_1\\
				\end{bmatrix}
				\begin{bmatrix}
					\boldsymbol{C}_1 & -\boldsymbol{S}_1 \\
					\boldsymbol{S}_1 & \boldsymbol{C}_1\\
				\end{bmatrix}
				\begin{bmatrix}
					\boldsymbol{W}_1 & \boldsymbol{0} \\
					\boldsymbol{0} & \boldsymbol{W}_1\\
				\end{bmatrix}
				\begin{bmatrix}
					\boldsymbol{D} & \boldsymbol{0} \\
					\boldsymbol{0} & \boldsymbol{D}^{-1}\\
				\end{bmatrix}
				\begin{bmatrix}
					\boldsymbol{W}_2 & \boldsymbol{0} \\
					\boldsymbol{0} & \boldsymbol{W}_2\\
				\end{bmatrix}
				\begin{bmatrix}
					\boldsymbol{C}_2 & -\boldsymbol{S}_2 \\
					\boldsymbol{S}_2 & \boldsymbol{C}_2\\
				\end{bmatrix}
				\begin{bmatrix}
					\boldsymbol{V}_2 & \boldsymbol{0} \\
					\boldsymbol{0} & \boldsymbol{V}_2\\
				\end{bmatrix},
			\end{equation}
			where $\boldsymbol{V}_i, \boldsymbol{W}_i \in U(n)$;
			$\boldsymbol{C}_i = \t{diag}(\cos \theta_{1,i},\ldots,\cos 
			\theta_{n,i}), 
			\boldsymbol{S}_i = \t{diag}(\sin \theta_{1,i},\ldots,\sin 
			\theta_{n,i})$,
			with $\theta_{k,i} \in \mathbb{R}$; and, 
			$\boldsymbol{D} = \t{diag}(e^{r_1},\ldots, e^{r_n})$ with $r_k \in 
			\mathbb{R}$.
			Without loss of generality, for each of the $\boldsymbol{V}_1,
			\boldsymbol{V}_2, \boldsymbol{W}_1$ matrices, $n$
			phases among their independent matrix coefficients can be fixed. 
		\end{theorem}
	\end{widetext}
	
	The physical import of this result is that
	any transfer function $\boldsymbol{M}[\omega]$ can be implemented
	as a succession of a $n$-mode frequency-dependent Mach-Zehnder,
	a set of frequency-dependent single mode squeezers, and another $n$-mode
	frequency-dependent Mach-Zehnder, as shown in 
	\cref{fig:Optical_decomposition}. 
	To realize a $n$-mode frequency dependent
	Mach-Zehnder, it suffices to apply the algorithm outlined in the
	frequency-independent case \cite{Reck1994, Clements:16} with
	frequency-dependent phase-shifters and beam-splitters, both of which can be 
	realized by detuning an optical cavity. 
	
	\Cref{thm:opt_decomp} is a generalization of two central results in 
	synthesis in two
	different fields of study. On the one hand, in quantum optics of 
	frequency-independent modes, 
	it is a generalization of the
	fact that a real symplectic transformation of
	$n$ modes can be decomposed into a set of $n$ single-mode squeezers 
	sandwiched
	between two $n$-mode Mach-Zehnder interferometers \cite{Bloch62, 
	Balian65,Arvind_1995,serafini23},
	where the latter can be implemented using a set of two-mode beam-splitters 
	and phase
	shifters \cite{Reck1994,Clements:16}.
	On the other hand, it is the natural generalization of the classical theory 
	of electrical network 
	synthesis \cite{Cauer58,Belev68,Youla15}; the subject of this theory is to 
	synthesize an electrical network
	whose response matches that of a specified ``realizable'' transfer 
	function. 
	The fundamental difficulty in that realm is to ensure good impedance match 
	between the various stages 
	whose cascade forms the full network; if not, succeeding stages load the 
	preceding ones. 
	The quantum analog of such loading is back-action \cite{Yurk91}. 
	So in a practical implementation, stringent control of extraneous 
	back-action, such as back-scatter
	in an optical implementation, will need to be exercised.
	
	\subsection{Related descriptions of linear quantum systems}
	\label{sec:SpH_related}
	
	The LTI transformation of the quantum noises can be equivalently
	described in terms of the creation/annihilation operators 
	$\boldsymbol{\hat{\mathfrak{a}}}(t) \eqdef 
	[\hat{a}_1(t),\ldots,\hat{a}_n(t),
	\hat{a}_1^\dagger(t),\ldots,\hat{a}_n^\dagger(t)]^\T$,
	where $\hat{a}_i(t) \eqdef (\hat{q}_i(t) + \ii \hat{p}_i(t))/\sqrt{2}$. 
	Since
	this transformation 
	$\boldsymbol{\hat{x}}\mapsto \boldsymbol{\hat{\mathfrak{a}}} = 
	\boldsymbol{P}^\dagger \boldsymbol{\hat{x}}$ 
	is implemented by the unitary matrix
	\begin{equation}
		\label{eq:P_matrix} \boldsymbol{P}  := 
		\frac{1}{\sqrt{2}}\begin{bmatrix}
			\boldsymbol{1}_n & \boldsymbol{1}_n\\
			-i \boldsymbol{1}_n & i \boldsymbol{1}_n\\
		\end{bmatrix},
	\end{equation}
	\noindent the LTI transformation 
	$\boldsymbol{\hat{x}}_\t{in}[\omega] \mapsto 
	\boldsymbol{\hat{x}}_\t{out}[\omega] 
	= \boldsymbol{M}[\omega] \boldsymbol{\hat{x}}_\t{in}[\omega]$, 
	induces the LTI transformation on creation/annihilation operators, 
	$\boldsymbol{\hat{\mathfrak{a}}}_\t{in}[\omega] \mapsto 
	\boldsymbol{\hat{\mathfrak{a}}}_\t{out}[\omega]
	= 
	(\boldsymbol{P}^\H\boldsymbol{M}[\omega]\boldsymbol{P})\boldsymbol{\hat{\mathfrak{a}}}_\t{in}[\omega]$.
	The former transformation preserves the CCR in \cref{eq:qwnCommutators}, 
	whereas the latter transformation
	preserves the CCR
	\begin{equation}\label{eq:CCRa}
		\left[\hat{\mathfrak{a}}_j[\omega],
		\hat{\mathfrak{a}}_k[\omega']^\dagger \right] = 2\pi \delta[\omega
		-\omega'] \ (\boldsymbol{I}_n)_{jk},
	\end{equation}
	where $\boldsymbol{I}_n \eqdef \boldsymbol{P}^\dagger \ii 
	\boldsymbol{J}_{2n} \boldsymbol{P} 
	= \t{diag}(\boldsymbol{1}_n, -\boldsymbol{1}_n)$. This motivates the 
	following definition.
	\begin{definition}\label{def:Upq}
		The indefinite unitary group of signature $(n,m)$ is the set
		\begin{equation}
			U(n,m)\eqdef
			\left\{\boldsymbol{A} \in \mathcal{M}_{n+m}(\mathbb{C})\  | \
			\boldsymbol{A}\, \boldsymbol{I}_{n,m}  \boldsymbol{A}^\H
			= \boldsymbol{I}_{n,m}\right\}
		\end{equation}
		where $\boldsymbol{I}_{n,m} = \t{diag}(\boldsymbol{1}_{n}, 
		-\boldsymbol{1}_{m})$;
		we denote $\boldsymbol{I}_n = \boldsymbol{I}_{n,n}$.
	\end{definition}
	
	The group $U(n,n)$ then describes CCR-preserving LTI transformations of the 
	creation/annihilation operators,
	and we have the group isomorphism $\SpH(2n) \cong U(n,n)$, realized by the 
	map 
	$\boldsymbol{M} \mapsto \boldsymbol{P}^\H\boldsymbol{M}\boldsymbol{P}$.
	
	The oft-studied real symplectic group 
	\cite{Arvind_1995,Braunstein05,Simon_1994,WeedPiran12,serafini23} 
	of order $2n$, $\t{Sp}(2n,\mathbb{R})$, 
	emerges from the conjugate symplectic order $\SpH(2n)$ in one of two 
	different ways.
	
	In the frequency-independent case, 
	$\boldsymbol{M}[\omega]=\boldsymbol{M}[0]$; 
	the symmetry property [\cref{eq:freq_domain_condition}] 
	$\boldsymbol{M}[\omega]^* = \boldsymbol{M}[-\omega]$ then 
	implies that $\boldsymbol{M} \eqdef \boldsymbol{M}[0]$ is real-valued. Thus 
	the conjugate symplectic condition 
	[\cref{eq:conjugate_symp_condition}] reduces to $\boldsymbol{MJM}^T = 
	\boldsymbol{J}$, i.e. 
	$\boldsymbol{M} \in \t{Sp}(2n,\mathbb{R})$ in the frequency-independent 
	case, and so
	$\t{Sp}(2n,\mathbb{R}) \subset \SpH(2n)$.
	
	More generally, the operators
	\begin{equation}\label{eq:xfmu} 
		\hat{x}_\mu \eqdef \int f_{\mu i}(t) \hat{x}_i(t) \dd t 
		= \int f_{\mu i}^* [\omega] \hat{x}_i[\omega] \frac{\dd \omega}{2\pi},
	\end{equation}
	defined by sampling the quantum noise with the real-valued ``window'' 
	functions $\{f_{\mu i}(t)\}$
	which satisfy $\int f_{\mu i}(t) f_{\nu j} \dd t = \delta_{\mu 
	i}\delta_{\nu j}$, are  
	canonically conjugate variables of a set of harmonic oscillators, i.e.
	$
	[\hat{x}_\mu, \hat{x}_\nu] = \ii J_{\mu \nu}.
	$ 
	
	A CCR-preserving LTI transformation of the quantum noises 
	$\{\hat{x}_i[\omega]\}$ induces a 
	CCR-preserving (i.e. canonical) linear transformation of $\{\hat{x}_\mu\}$. 
	The latter forms the
	group $\t{Sp}(2n,\mathbb{R})$; but since the map \cref{eq:xfmu} is 
	many-to-one, it realizes 
	the inclusion $\t{Sp}(2n,\mathbb{R}) \subset \SpH(2n)$. 
	The physical interpretation of sampling is useful in defining
	harmonic oscillator modes out of propagating quantum optical 
	fields \cite{Blow_Loudon,Fabre_2020,RoyDev16}; we will revisit this aspect
	in \cref{sec:realization_SDM} in the context of characterizing a quantum 
	LTI system.

	The reason that the conjugate symplectic group is fundamentally distinct 
	from the
	real symplectic group is that the former describes linear transformations
	of the \emph{non-hermitian} operators $\boldsymbol{\hat x}[\omega] = 
	[\boldsymbol{\hat{q}}[\omega], 
	\boldsymbol{\hat{p}}[\omega]]^T$.
	This can be ameliorated by defining, for $\omega \geq 0$, the set of $4n$ 
	\emph{hermitian} operators
	\begin{equation}
		\boldsymbol{\hat{\mathfrak{X}}}[\omega] = 
		\frac{1}{\sqrt{2}}\begin{bmatrix} 
			\boldsymbol{\hat{x}}[+\omega] + \boldsymbol{\hat{x}}[-\omega] \\
			-\ii (\boldsymbol{\hat{x}}[+\omega] - \boldsymbol{\hat{x}}[-\omega])
		\end{bmatrix} =
		\mathbb{P} \begin{bmatrix} \boldsymbol{\hat{x}}[+\omega] \\ 
			\boldsymbol{\hat{x}}[-\omega] \end{bmatrix};
	\end{equation}
	here $\mathbb{P}$ is the $4n-$dimensional unitary matrix that performs the 
	required linear transformation.
	Partitioned into the form $\boldsymbol{\hat{\mathfrak{X}}}[\omega] = 
	[\boldsymbol{\hat{x}}^\mathcal{A} [\omega], 
	\boldsymbol{\hat{x}}^\mathcal{B} [\omega]]$, these are precisely the 
	``two-photon quadratures'' in the Caves-Schumaker 
	formalism 
	\cite{Caves85,Schumaker85,DanKhal12,Branford2018,Gefen2024,Gardner_2024} 
	Importantly, $\hat{x}_i^{\mathcal{A},\mathcal{B}} [\omega]$ are all 
	hermitian and are canonically conjugate of each other: 
	$[\hat{q}_j^{\mathcal{A}}[\omega],\hat{p}_k^\mathcal{A}[\omega']] = 2\pi 
	\ii 
	\delta_{jk} 
	\delta[\omega-\omega']$ (similarly for the $\mathcal{B}$ quadratures). 
	Thus, 
	the set of 
	CCR-preserving linear transformations acting on 
	$\boldsymbol{\hat{\mathfrak{X}}}[\omega]$ is the real symplectic group 
	$\t{Sp}(4n,\mathbb{R})$ \textit{for $2n$ modes}. However, such
	transformations generally mix $\boldsymbol{\hat{x}}[+\omega]$ and 
	$\boldsymbol{\hat{x}}[-\omega]$, indicating
	that they cannot arise from LTI transformations of 
	$\boldsymbol{\hat{x}}[\omega]$.
	Indeed the LTI transformation $\boldsymbol{\hat{x}}[\omega] \mapsto  
	\boldsymbol{M}[\omega]\boldsymbol{\hat{x}}[\omega]$ --- the totality of 
	which form the conjugate symplectic 
	group $\SpH(2n)$ --- induces the transformation 
	$\boldsymbol{\hat{\mathfrak{X}}}[\omega] \mapsto \mathbb{P} \,
	\t{diag}(\boldsymbol{M}[\omega],\boldsymbol{M}[-\omega])\mathbb{P}^\H\boldsymbol{\hat{\mathfrak{X}}}[\omega]$,
	which is clearly a
	subgroup of $\t{Sp}(4n,\mathbb{R})$ (see 
	Section IV of SI for proofs). Thus follows the 
	inclusion $\SpH(2n)\subset \t{Sp}(4n,\mathbb{R})$.
	
	In sum, the conjugate symplectic group of order $2n$ --- describing the LTI 
	transformations of
	quantum white noise --- encompasses the set of linear canonical 
	transformations of harmonic oscillator modes
	--- the subject of Gaussian quantum information studies --- and  is 
	included in the set of 
	linear ``two-photon'' transformations, i.e.
	\begin{equation}
		\t{Sp}(2n, \mathbb{R}) \subset \SpH(2n) \cong U(n,n) 
		\subset \t{Sp}(4n,\mathbb{R}).
	\end{equation}
	
	\section{Uncertainty principles for quantum LTI 
	systems}\label{sec:uncertainty}
	
	Given the black-box transfer matrix $\boldsymbol{M}[\omega]$ of a quantum 
	LTI system  ---
	either closed, in which case $\boldsymbol{M} \in \SpH$, or open, in which 
	case hidden noise
	modes are present --- the unavoidable implication of quantum mechanics is 
	that the outputs
	feature a certain minimum noise. In the following, we elucidate the 
	structure 
	and trade-off among the uncertainties in the quantum noises.
	
	We assume that the quantum noises are weak-stationary, i.e. 
	$\avg*{\hat{x}_i(t)\hat{x}_j(t')} =
	\avg*{\hat{x}_i(t-t')\hat{x}_j(0)}$ for all quantum noises and times 
	$t,t'$. 
	This assumption is consistent in the sense that an LTI map preserves the 
	weak-stationary property of inputs.
	Then the fluctuations
	in the $2n$ quantum noises $\hat{\boldsymbol{x}}(t)$
	are described by the (symmetrized) spectral density matrix (SDM) 
	$\bar{\boldsymbol{S}}[\omega]$,
	with elements
	\begin{equation}
		\bar{S}_{ij}[\omega]\times 2\pi \delta[0] =
		\frac{1}{2}\left\langle\acomm{\hat{x}_i[\omega]}{\hat{x}_j[\omega]^\dagger}\right\rangle,
	\end{equation}
	where $\acomm{\cdot}{\cdot}$ is the anti-commutator; here, and henceforth, 
	we assume without loss
	of generality that $\avg{\hat{x}_i} = 0$.
	The SDM is complex-valued, has the symmetry
	$\bar{\boldsymbol{S}}[\omega]^* = \bar{\boldsymbol{S}}[-\omega]$, and it is 
	hermitian
	positive semi-definite (HPSD), as shown in \cref{thm:SDM_SDP} of 
	\cref{app:notations}. 
	
	
	In the frequency-independent case, the noise properties are encoded in the 
	covariance matrix
	$\boldsymbol{V} \eqdef \bar{\boldsymbol{S}}[0]$. It is real-valued, 
	symmetric, and
	positive semi-definite.
	In this setting, the uncertainty principle states that \cite{Simon_1994}
	$\boldsymbol{V} + \ii\boldsymbol{J}/2$ is hermitian positive semi-definite, 
	i.e.
	\begin{equation}\label{eq:Vunc}
		\boldsymbol{V} + \frac{\ii}{2}\boldsymbol{J} \geq 0.
	\end{equation}
	In particular, for a single mode, it yields the inequality
	\begin{equation}\label{eq:real_up}
		V_{qq}V_{pp} - V_{qp}^2 \geq \frac{1}{4}.
	\end{equation}
	
	\subsection{Conjugate symplectic uncertainty principle and invariant 
	reduction of SDM}
	
	In the frequency-dependent case we can show (see SI 
	Section V) that the
	uncertainty principle can be stated as follows.
	
	\begin{theorem}\label{thm:up_nmode}
		Let $\SDMO{\omega}$ be the SDM of $n$ weak-stationary quantum noises. 
		For any $\omega \in
		\mathbb{R}$, the matrix $\SDMO{\omega} + \ii\boldsymbol{J}/2$ is
		hermitian, positive semi-definite, i.e.
		\begin{equation}
			\SDMO{\omega} + \frac{\ii}{2}\boldsymbol{J} \geq 0.
		\end{equation}
	\end{theorem}
	
	Despite the superficial similarity to the frequency-dependent case 
	[\cref{eq:Vunc}], the content of this
	inequality is subtly different. In particular, since the SDM is hermitian 
	(and not symmetric, like the covariance
	matrix), even for a single quantum noise $\{\hat{q}(t),\hat{p}(t)\}$, 
	\cref{thm:up_nmode} implies 
	\begin{equation}\label{eq:tightUncertainty}
		\bar{S}_{qq}[\omega] \bar{S}_{pp}[\omega] -
		|\bar{S}_{qp}[\omega]|^2 \geq \frac{1}{4} +
		|\Im\left(\bar{S}_{qp}[\omega]\right)|,
	\end{equation}
	which is tighter compared to the frequency-independent case 
	[\cref{eq:real_up}] due to the imaginary cross
	correlation term.
	
	This qualitatively novel feature turns out to be fundamental in the
	phenomenology of frequency-dependent quantum LTI systems. 
	In order to characterize the origin and structure of the imaginary 
	cross-correlation, 
	we seek a convenient normal form of the SDM. Since the LTI transformation
	$\boldsymbol{\hat{x}}[\omega]\mapsto \boldsymbol{M}[\omega] 
	\boldsymbol{\hat{x}}[\omega]$
	induces the transformation on the SDM $\SDMO{\omega} \mapsto 
	\boldsymbol{M}[\omega] \SDMO{\omega} 
	\boldsymbol{M}[\omega]^\dagger$, our interest is limited to a normal form 
	from which every 
	SDM can be generated by an LTI transformation. The following theorem 
	achieves precisely
	such a characterization (see SI Section VI for 
	proof).
	
	\begin{theorem}
		\label{thm:Williamson_SDM}
		Let $\SDM$ be the SDM of an $n$ weak-stationary quantum noises. For
		any frequency $\omega\in \mathbb{R}$, there exists
		\begin{itemize}
			\item $\boldsymbol{M}[\omega]\in \SpH(2n)$  such that
			$\boldsymbol{M}[-\omega] = \boldsymbol{M}[\omega]^*$;
			\item a diagonal matrix $\boldsymbol{\sigma}[\omega] =
			\textnormal{diag}(\sigma_1[\omega],\dots,\sigma_n[\omega])$ with
			$\sigma_j $ a positive even function for any $j\in \{1,\dots, n\}$;
			\item a diagonal matrix $\boldsymbol{\Delta}[\omega] =
			\textnormal{diag}(\Delta_1[\omega],\dots,\Delta_n[\omega])$ with
			$\Delta_j$ an odd function for any $j\in \{1,\dots, n\}$, satisfying
			\begin{equation}\label{eq:delta_1/2}
				\sigma_j[\omega] \geq  \frac{1}{2} + |\Delta_j[\omega]|;
			\end{equation}
		\end{itemize}
		such that
		\begin{equation}\label{eq:conjugate_williamson}
			\SDMO{\omega} = \boldsymbol{M}[\omega] 
			\begin{bmatrix}
				\boldsymbol{\sigma}[\omega] & +\ii \boldsymbol{\Delta}[\omega]\\
				- \ii \boldsymbol{\Delta}[\omega] &
				\boldsymbol{\sigma}[\omega]\\
			\end{bmatrix}
			\boldsymbol{M}[\omega]^\H.
		\end{equation}
		The elements $\{\boldsymbol{\Sigma},\boldsymbol{\Delta}\}$ are given by
		\begin{equation} \label{eq:sigma_j_delta_j_def}
			\qquad \sigma_j = \frac{\mu_j - \mu_{n+j}}{2}>0,  \qquad \Delta_j = 
			\frac{\mu_j + \mu_{n+j}}{2},
		\end{equation}
		where $\mu_1\geq\cdots \geq \mu_n > 0 > \mu_{n+1}\geq \cdots \geq 
		\mu_{2n}$
		are the eigenvalues (counting degeneracy) of the matrix $\ii 
		\boldsymbol{J\bar{S}}$
		in decreasing order.
	\end{theorem}
	
	This generalizes Williamson's theorem \cite{Williamson36} for the real 
	symplectic group.
	To wit, we know that (setting $\omega = 0$), $\boldsymbol{M}[0]\in 
	\t{Sp}(2n,\mathbb{R})$, 
	and \cref{thm:Williamson_SDM} asserts that, $\boldsymbol{V} = 
	\boldsymbol{\bar{S}}[0] = \boldsymbol{M}\,
	\t{diag}(\boldsymbol{\Sigma}, \boldsymbol{\Sigma})\, 
	\boldsymbol{M}^\dagger$.
	(Here, $\boldsymbol{\Delta}$ vanished due to its odd symmetry.)
	That is, any real covariance matrix
	can be generated by a real symplectic transformation of a diagonal 
	covariance matrix.
	It can be shown that there exists an $n-$mode thermal state
	with average occupation \cite{Simon_1994,WeedPiran12,serafini23} 
	$\avg*{\hat{n}_j} = \Sigma_j -\tfrac{1}{2}$.
	That is, any covariance matrix of an $n$ frequency-independent modes can be 
	prepared by a
	real symplectic transformation of a suitable set of $n$ thermal states. 
	
	\cref{thm:Williamson_SDM} states that the frequency-dependent case is 
	qualitatively different.
	First, a general SDM results from a conjugate symplectic transformation
	of the irreducible (non-diagonal) SDM $[\boldsymbol{\Sigma}, +\ii 
	\boldsymbol{\Delta}; 
	-\ii \boldsymbol{\Delta}, \boldsymbol{\Sigma}]$.
	The imaginary off-diagonal term here is precisely the origin of the
	imaginary cross-correlation in the uncertainty bound in 
	\cref{eq:tightUncertainty}; in fact, \cref{eq:delta_1/2} is equivalent to 
	\cref{eq:tightUncertainty}.
	Thus, complete characterization of the outputs of such systems need to be 
	sensitive to the 
	imaginary cross-correlation $\ii \boldsymbol{\Delta}$; 
	\cref{sec:realization_SDM} illustrates such 
	measurement schemes.
	Second, the pair $\{\boldsymbol{\Sigma},\boldsymbol{\Delta}\}$ captures the 
	features of the
	quantum noises that are invariant to conjugate symplectic transformations.
	That is, the output of a \emph{closed} quantum LTI system will have the 
	same 
	$\{\boldsymbol{\Sigma},\boldsymbol{\Delta}\}$ as its input.
	Therefore these features can only arise in the first place from an open 
	quantum LTI system.
	
	\subsection{Nature and origin of the conjugate symplectic invariants}
	
	An open system can be modeled starting from a closed system with a 
	conjugate symplectic
	matrix $\boldsymbol{M}^\t{ext}$, explicitly partitioning it into its 
	accessible and inaccessible parts 
	[\cref{eq:Mext}] $\boldsymbol{M}^\t{ext} = 
	[\boldsymbol{G}, \boldsymbol{N}; \boldsymbol{K},\boldsymbol{L}]$, then 
	using the generalized Williamson
	decomposition to write its total output SDM
	\begin{equation}
		\boldsymbol{\bar{S}}[\omega] = 
		\begin{bmatrix} \boldsymbol{G} & \boldsymbol{N} \\ \boldsymbol{K} & 
		\boldsymbol{L}\end{bmatrix}
		\begin{bmatrix} \boldsymbol{\Sigma} & +\ii\boldsymbol{\Delta} 
			\\ -\ii \boldsymbol{\Delta} & \boldsymbol{\Sigma}\end{bmatrix}
		\begin{bmatrix} \boldsymbol{G} & \boldsymbol{N} \\ \boldsymbol{K} & 
		\boldsymbol{L}\end{bmatrix}^\dagger,
	\end{equation}
	and omitting all but the accessible output (as in 
	\cref{fig:quantumLti}(c)). 
	The SDM of this accessible output is the top left corner of the above 
	matrix (independent of 
	$\boldsymbol{K}, \boldsymbol{L}$; we also assume for simplicity that the 
	subsystem
	dimensions are compatible as follows):
	\begin{equation}\label{eq:SoutMN}
		\begin{split}
			\boldsymbol{\bar{S}}^\t{out} = &(\boldsymbol{G\Sigma G}^\dagger + 
			\boldsymbol{N\Sigma N}^\dagger)+ \ii (\boldsymbol{G\Delta 
			N}^\dagger - 
			\boldsymbol{N\Delta G}^\dagger).\\
		\end{split}
	\end{equation}
	The first term can be interpreted as noise at the outputs (accessible or 
	inaccessible) of the system 
	transferred from the respective inputs. 
	By contrast, the second term represents the output arising from 
	correlations between the accessible and inaccessible inputs. 
	
	The precise interpretation of these matrices can be elucidated by 
	transforming the irreducible SDM in terms of the creation/annihilation
	operators $\boldsymbol{\hat{\mathfrak{a}}}$:
	\begin{equation}
		\begin{bmatrix}
			\boldsymbol{\sigma}[\omega] & +\ii \boldsymbol{\Delta}[\omega]\\
			- \ii \boldsymbol{\Delta}[\omega] &
			\boldsymbol{\sigma}[\omega]\\
		\end{bmatrix} \mapsto 
		\begin{bmatrix}
			\boldsymbol{\sigma}[\omega] - \boldsymbol{\Delta}[\omega] & 
			\boldsymbol{0}\\
			\boldsymbol{0} & \boldsymbol{\sigma}[\omega] + 
			\boldsymbol{\Delta}[\omega] \\
		\end{bmatrix},
	\end{equation}
	\noindent (see \cref{eq:simplified_SDM_a}) and identifying the latter with 
	the 
	generic SDM in terms of the creation/annihilation
	operators, whose diagonal elements are $\avg*{\hat{a}_j[\mp \omega]^\dagger 
	\hat{a}_j[\mp \omega]} +1 = 
	\avg*{\hat{n}_j[\mp \omega]}+1$. This gives, 
	$\avg*{\hat{n}_j[\pm\omega]} = \sigma_j[\omega] \pm\Delta_j[\omega] 
	- \frac{1}{2}$, or equivalently
	\begin{align}\label{eq:sigma_interpretation}
		\sigma_j[\omega] &= \frac{\langle \hat{n}_j[+\omega]\rangle +\langle 
			\hat{n}_j[-\omega]\rangle}{2} + \frac{1}{2}\\
		\Delta_j[\omega] &= \frac{\langle \hat{n}_j[+\omega]\rangle -\langle 
			\hat{n}_j[-\omega]\rangle}{2} \label{eq:delta_interpretation}
	\end{align}
	That is, $\sigma_j[\omega]$ represents the symmetrized 
	thermal occupation, while $\Delta_j[\omega]$ is the 
	sideband asymmetry in the average quantum number. 
	
	\begin{figure}[t!]
		\centering
		\includegraphics[width=\linewidth]{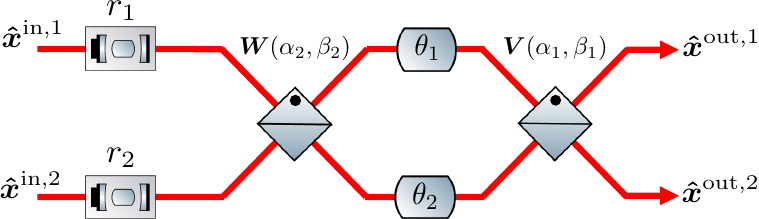}
		\caption{Two-mode general symplectic transformation acting on vacuum. 
			Appropriately choosing the frequency-dependent coefficients
			$\alpha_1,\alpha_2,\beta_1,\beta_2,\theta_1$ and $\theta_2$ enables 
			the 
			generation of the $\{\sigma, \Delta\}$ form of 
			\cref{eq:sigma_delta_SDM}.}
		\label{fig:two-mode-vacuum-scheme}
	\end{figure}
	
	A concrete scheme to generate sideband asymmetry
	in the output of an open system can be constructed by employing the
	optical decomposition of the conjugate symplectic group 
	(\cref{thm:opt_decomp}). 
	The simplest possibility is a closed system that acts on a pair of quantum 
	noises, 
	followed by elimination of one of its outputs. For example, consider the 
	system depicted in
	\cref{fig:two-mode-vacuum-scheme}, which is represented by
	the $\SpH(4)$ transfer matrix
	\begin{equation}
		\boldsymbol{M} = \begin{bmatrix}
			\boldsymbol{D} & \boldsymbol{0}\\
			\boldsymbol{0} & \boldsymbol{D}^{-1}
		\end{bmatrix}
		\begin{bmatrix}
			\boldsymbol{W}_2 & \boldsymbol{0}\\
			\boldsymbol{0} & \boldsymbol{W}_2
		\end{bmatrix}
		\begin{bmatrix}
			\boldsymbol{C} & -\boldsymbol{S} \\
			\boldsymbol{S} & \boldsymbol{C}\\
		\end{bmatrix}
		\begin{bmatrix}
			\boldsymbol{W}_1 & \boldsymbol{0}\\
			\boldsymbol{0} & \boldsymbol{W}_1
		\end{bmatrix}
	\end{equation}
	where $\boldsymbol{D} = \text{diag}(e^{r_1}, e^{r_2})$, 
	$\boldsymbol{C} = \text{diag}(\cos\theta_1, \cos\theta_2)$, 
	$\boldsymbol{S} = \text{diag}(\sin\theta_1, \sin\theta_2)$, 
	and $\boldsymbol{W}_i = [\cos \alpha_i, -e^{i\beta_i} \sin \alpha_i; 
	e^{-i \beta_i} \sin \alpha_i, \cos \alpha_i]$
	with the choice $\alpha_1 = \alpha_2 = \pi/4, \beta_1 = \pi/2, \beta_2 = 
	0, \theta_1 = \pi/2, \theta_2 = 0$.
	Let this system be fed with vacuum inputs, i.e. 
	$\boldsymbol{\bar{S}}^\t{in} =\boldsymbol{1}/2$.
	Then the SDM of one of its outputs (i.e. after eliminating the other) is 
	\begin{equation}\label{eq:sigma_delta_SDM}
		\boldsymbol{\bar{S}}^\text{out,2} = \begin{bmatrix}
			\Sigma & \ii \Delta \\
			-\ii \Delta & \Sigma
		\end{bmatrix}
	\end{equation}
	where 
	\begin{align}
		\Sigma &= \frac{1}{2} \left(\cosh^2r_2 + \cosh^2r_1\right)\\
		\Delta &= 
		\frac{1}{2} \left( \sinh^2r_2 - \sinh^2r_1\right)
	\end{align}
	
	That is, the outputs of a pair of squeezers, passed through an
	interferometer, produces in one of its arms, after having eliminated the 
	other,
	the irreducible SDM that appears in the generalized Williamson 
	decomposition.
	This scheme can be naturally generalized to generate a 
	$n-$mode irreducible SDM by 
	starting with a $2n-$mode vacuum and applying the scheme for the $n$ pairs 
	of 
	modes, then tracing out on half of them.

	\subsection{Tighter uncertainty bound for open LTI systems}
	
	The uncertainty relations derived above 
	[\cref{eq:tightUncertainty,eq:delta_1/2}] apply when the 
	system is closed, and therefore described by a conjugate symplectic 
	transfer matrix. Quite often, 
	as discussed in \cref{subsec:LTIdilated}, only an open subsystem is 
	accessible.
	This subsystem is then described by an input-output relation 
	[\cref{eq:quantumLtiOp}] 
	\begin{equation}\label{eq:quantumLtiOp1}
		\boldsymbol{\hat{x}}^\text{out}[\omega] = \boldsymbol{G}[\omega]
		\boldsymbol{\hat{x}}^\text{in}[\omega] + \boldsymbol{N}[\omega]
		\boldsymbol{\hat{x}}^\text{N}[\omega],
	\end{equation}
	where $\boldsymbol{G}$ is usually known, and $\boldsymbol{N}$ 
	is reconstructed from it (see \cref{subsec:quantumNoiseInLtiSystems}); 
	neither 
	of them need be conjugate symplectic.
	The theorem below bounds the uncertainty of the accessible outputs when the 
	inaccessible inputs cannot be quantum engineered, thus yielding a stricter 
	result than the ones derived above for closed systems (see SI 
	Section VII for a 
	proof).
	\begin{theorem}\label{thm:tight_open}
		Assuming that the noise modes (i.e. inaccessible inputs) are 
		uncorrelated 
		with each other and with the input modes, the output modes' spectral 
		densities satisfy, for all $i\in\{1,\dots, n\}$
		\begin{equation}\label{eq:generalizedUncertainty}
			\sqrt{\bar{S}^\text{out}_{q_i q_i} \bar{S}^\text{out}_{p_i p_i}} 
			\geq \sqrt{ \det {\sum_{j,k}} \boldsymbol{G}_{ij} 
				\boldsymbol{\bar{S}}_{jk}^\text{in} \boldsymbol{G}_{ki}^\dagger 
			} 
			+ \frac{1}{2}{\sum_j} \left|\det \boldsymbol{N}_{ij} 
			\right|.
		\end{equation}
		
		\noindent where the matrices on the right hand side are defined as: for 
		$\boldsymbol{A} \in \{\boldsymbol{G},\boldsymbol{\bar{S}}^\text{in}, 
		\boldsymbol{N}\}$,
		\begin{equation}\label{eq:oneModeSdmBlock}
			\boldsymbol{{A}}_{jk} \eqdef 
			\begin{bmatrix} {A}_{q_j q_k} & {A}_{q_j p_k} \\
				{A}_{p_j q_k} & {A}_{p_j p_k} \end{bmatrix}.
		\end{equation}
	\end{theorem}
	
	Note that this inequality is tight when $\bar{S}_{q_ip_i}=0$, and all of 
	the 
	matrices $\boldsymbol{N}_{ij}\boldsymbol{N}_{ji}^\dagger$ are proportional 
	to 
	each other and to ${\textstyle\sum_{j,k}} \boldsymbol{G}_{ij} 
	\boldsymbol{\bar{S}}_{jk}^\text{in} \boldsymbol{G}_{ki}^\dagger$ (if this 
	term 
	is non-zero). 
	The former is true if, for instance, all of these transfer matrices treat 
	the 
	amplitude and phase quadratures symmetrically.
	Since the first term on the right-hand-side of 
	\cref{eq:generalizedUncertainty} 
	is positive semi-definite, 
	the looser form, 
	\begin{equation}\label{eq:generalizedUncertaintyNoiseOnly}
		\begin{split}
			\sqrt{\bar{S}^\text{out}_{q_i q_i} \bar{S}^\text{out}_{p_i p_i}} 
			\geq 
			\frac{1}{2} \sum_j 
			\left|\det \boldsymbol{N}_{ij} \right|
		\end{split}
	\end{equation}
	also holds, which is still tighter than \cref{thm:up_nmode}.
	
	\section{SDM tomography}\label{sec:realization_SDM}
	
	\begin{figure}[t!]
		\centering
		\includegraphics[width=0.9\columnwidth]{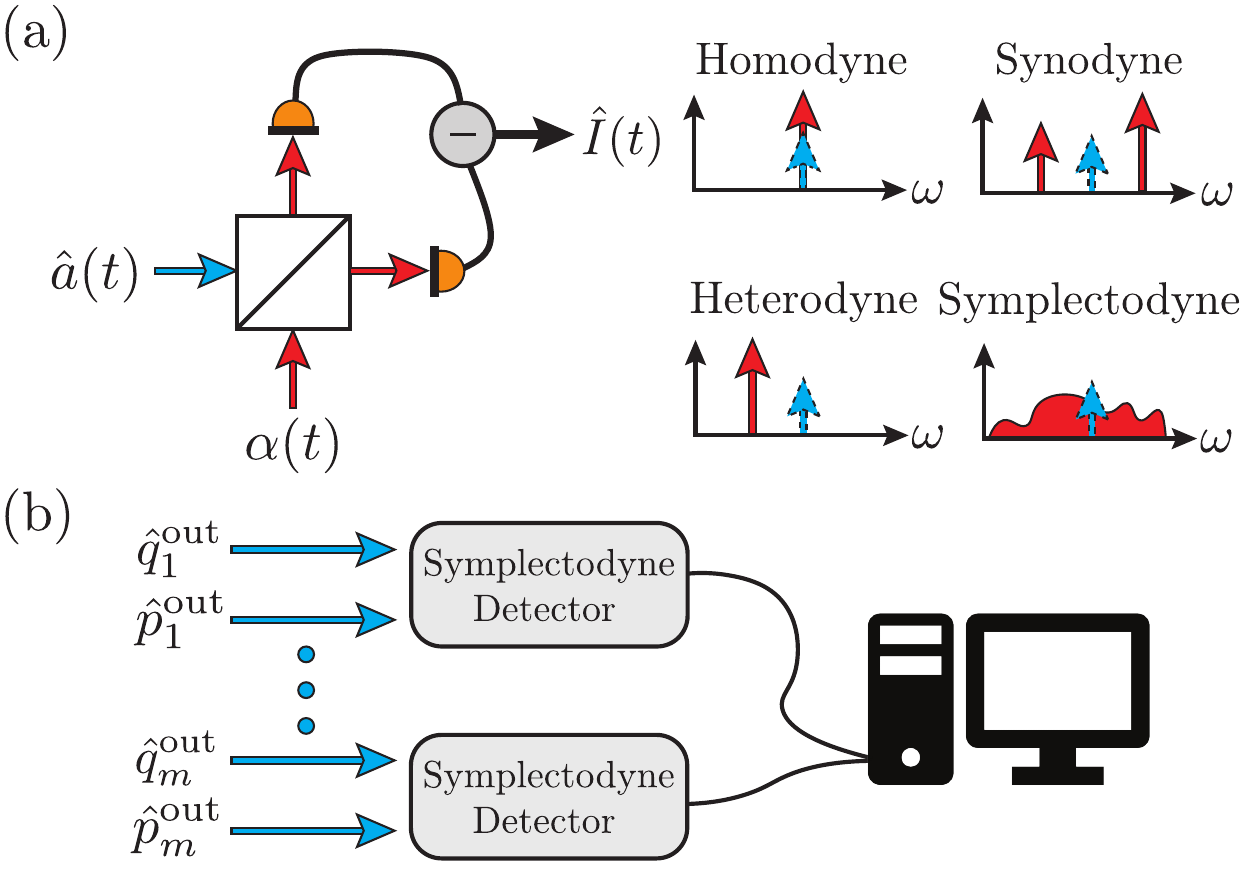}
		\caption{\textit{Symplectodyne detector.} (a) Shows the implementation 
		of a 
			single-mode symplectodyne detector. The local oscillator 
			$\alpha(t)$ with an 
			arbitrary waveform is coupled to a signal field $\hat{a}(t)$ with a 
			balanced coupler. The spectrum of the differential photocurrent 
			$\hat{I}(t)$ 
			is used to recover the spectral density matrix of the 
			signal. Depending on the local oscillator spectrum, symplectodyne 
			detection subsumes 
			homodyne, heterodyne, or synodyne detection. The 
			relationship between the local oscillator tone (red arrows) and 
			carrier 
			frequency (dashed blue arrows) is indicated for each detection 
			scheme. 
			(b) Multimode SDM tomography can be performed by subjecting the 
			output fields of a quantum 
			LTI system to symplectodyne detection and 
			correlating their output photocurrents. }
		\label{fig:symplectodyne}
	\end{figure}
	
	The central role of the SDM in characterizing the noise performance
	of quantum LTI systems demands the availability of techniques to 
	experimentally reconstruct it
	given access to the output ports of a quantum LTI system. Surprisingly, as 
	we will show, conventional 
	homodyne (or heterodyne) measurements
	of the output noises are not sufficient to reconstruct the SDM. This is a 
	manifestation of the imaginary cross-correlations
	encoded in the SDM, and its retrieval calls for a modification of the 
	conventional homodyne scheme.
	We now describe a scheme --- called \emph{symplectodyne} measurement --- 
	that can overcome the 
	incompleteness inherent in conventional measurements.
	
	\subsection{Single mode case}
	\label{sec:synodyne}
	
	To illustrate the principle of symplectodyne measurement, we first consider 
	the simple problem of SDM 
	tomography of a single mode.
	The tomographic reconstruction of its SDM 
	entails the estimation of the elements of the hermitian matrix
	\begin{equation*}
		\SDMO{\omega_0} = \begin{bmatrix}
			\bar{S}_{qq}[\omega_0] & \bar{S}_{qp}[\omega_0] \\
			\bar{S}_{qp}[\omega_0]^* & \bar{S}_{pp}[\omega_0]
		\end{bmatrix}
	\end{equation*}
	at each frequency $\omega_0$; i.e.
	four real parameters (two diagonal terms,
	the real and imaginary parts of the cross-correlation) per frequency.
	
	Consider the scheme shown in \cref{fig:symplectodyne}(a),
	wherein a balanced photodetector is fed with the equal superposition of a 
	local
	oscillator with classical complex-valued amplitude waveform $\alpha(t)$ and 
	the ladder operator
	$\hat{a}(t)$ of the single-mode ``signal'' of interest.
	The linearized photocurrent emitted by the detector is (up to a 
	multiplicative constant, 
	taken to be $1/\sqrt{2}$ for convenience)
	\begin{equation}\label{eq:symplectodyne_photocurrent}
		\hat{I}(t) \eqdef \frac{\alpha^*(t) \hat{a}(t) + 
			\alpha(t) \hat{a}^\dagger(t)}{\sqrt{2}} = 
		\boldsymbol{\alpha}(t)^\dagger\,  \boldsymbol{\hat{x}}(t);
	\end{equation}
	here, we have defined the two (real-valued) 
	local oscillator quadratures $\alpha_q(t) = [\alpha(t) + \alpha^*(t)]/2$ 
	and $\alpha_p(t) = \ii[\alpha(t)^* - \alpha(t)]/2$ arranged into the vector 
	$\boldsymbol{\alpha}(t) = [\alpha_q(t), \alpha_p(t)]^T$. 
	This photocurrent operator is effectively a classical stochastic process 
	if $[\hat{I}(t),\hat{I}(t')] =0$ for all $t,t'$ \cite{Bela94}; this 
	condition boils
	down to $J_{ij}\alpha_i(t)\alpha_j(t) = 0$, which is always satisfied on 
	account of the
	anti-symmetry of $\boldsymbol{J}$. Thus, the symplectic matrix plays an 
	essential
	role, hence the name ``symplectodyne'' measurement.

	The spectrum of the symplectodyne photocurrent is 
	\begin{equation}\label{eq:symplectodyne_general_spectrum}
		\bar{S}_{II}[\Omega] \times 2\pi\delta[0] = 
		\int_{-\infty}^{+\infty}\frac{\dd\Omega}{2\pi} 
		\boldsymbol{\alpha}[\Omega]^\dagger \, \SDMO{\omega + \Omega} 
		\boldsymbol{\alpha}[\Omega]
	\end{equation}
	Choice of the spectral content of the LO, i.e. the functional form 
	$\boldsymbol{\alpha}[\Omega]$,
	subsumes the homodyne, heterodyne, or synodyne
	detection \cite{Buchmann_2016} strategies. However, neither homodyne or 
	heterodyne
	photocurrent spectra contain the full information contained in the SDM of 
	the signal.
	
	To wit, the homodyne case corresponds to the choice 
	$\boldsymbol{\alpha}[\omega] = 
	2\pi\delta[\omega] |\alpha_0| \times 
	\left[\cos\theta, 
	\sin\theta\right]^T$ with $\theta = \arg\alpha_0$. Then 
	\cref{eq:symplectodyne_general_spectrum} reduces to 
	\begin{equation}\label{eq:HD_symmetric_form}
		\bar{S}_{II}^\t{hom}[\omega] = 
		\boldsymbol{\alpha_0}^\T \Re\left(\SDMO{\omega}\right) 
		\boldsymbol{\alpha_0}
	\end{equation}
	where $\boldsymbol{\alpha_0} = 
	[\Re(\alpha_0), \Im(\alpha_0)]^\T \in \mathbb{R}^2$.
	Note that the photocurrent is only sensitive to the real
	part of the off-diagonal elements ($\bar{S}_{qp}$) of the SDM; i.e. 
	homodyne detection is blind to the
	imaginary cross-correlations \cite{Barb_PRA,Gouz23}, which can hide quantum 
	resources detailed in \cref{sec:dephasing_experiment}.
	Heterodyne detection corresponds to the choice 
	$\alpha[\omega] = 2\pi \alpha_0\delta[\omega-\omega_0]$, where 
	$\alpha_0\in\mathbb{C}$, so that
	\cref{eq:symplectodyne_general_spectrum} reduces to
	\begin{align}
		\bar{S}_{II}^\t{het}[\omega] = \frac{|\alpha_0|^2}{4} \big[& 
		\bar{S}_{qq}[\omega_0 + \omega] +\bar{S}_{qq}[\omega_0 - \omega] 
		\nonumber \\
		& + \bar{S}_{pp}[\omega_0 + \omega] +\bar{S}_{pp}[\omega_0 - \omega] 
		\nonumber \\
		& + 2\Im(\bar{S}_{qp}[\omega_0 + \omega] +\bar{S}_{qp}[\omega_0 - 
		\omega])\big].
	\end{align}
	Contrary to homodyne detection the heterodyne spectrum 
	reveals the quantum noise at both positive and negative offsets around 
	$\omega_0$,
	but it is incomplete in the sense
	that it isn't sensitive to the real part of the cross-correlation.
	
	Complete information in the signal SDM can be captured using a 
	two-tone local oscillator symmetric about the carrier 
	frequency, i.e. the choice
	$\alpha[\omega] = \sqrt{2}\pi \left(\alpha_+\delta[\omega-\omega_0] + 
	\alpha_- \delta[\omega+\omega_0]\right)$, where $\alpha_+, 
	\alpha_-\in\mathbb{C}$. 
	This scheme is a form of quadrature amplitude demodulation (see 
	Supplementary Information 
	Section VIII), and has apparently been 
	re-discovered under the
	guise of ``synodyne'' detection \cite{Buchmann_2016}. 
	With this choice of LO, \cref{eq:symplectodyne_general_spectrum} at $\omega 
	=0$ reduces to
	\begin{equation}\label{eq:SDM_tomo_quadratic_form}
		\bar{S}_{II}[0] = \boldsymbol{\alpha_0}^\dagger 
		\SDMO{\omega_0}\boldsymbol{\alpha_0} \defeq Q(\alpha_{q0}, \alpha_{p0}),
	\end{equation}
	a hermitian form $Q(\cdot,\cdot)$ of the elements of the vector
	\begin{equation}\label{eq:synondyne_quadrature_expression}
		\boldsymbol{\alpha_0} = \frac{1}{2} \begin{bmatrix}
			\alpha_+ + \alpha_-^*\\
			\ii\left(\alpha_-^* -\alpha_+\right)
		\end{bmatrix} \defeq \begin{bmatrix}
			\alpha_{q0}\\
			\alpha_{p0}
		\end{bmatrix}\in\mathbb{C}^2.
	\end{equation}
	The SDM $\SDMO{\omega_0}$ can be reconstructed from
	photocurrent measurements by suitable choices of 
	$\boldsymbol{\alpha_0}$.
	Indeed, its diagonal entries are given by 
	$\bar{S}_{qq}[\omega_0] = Q(\alpha_{q0}, 0) /|\alpha_{q0}|^2 $ and 
	$\bar{S}_{pp}[\omega_0] = Q(0, \alpha_{p0})/ |\alpha_{p0}|^2 $, while the
	real and imaginary parts of the cross-correlation are given by (tuning the 
	relative phase between the $q$ and $p$ quadratures of the modulated LO):
	\begin{subequations}
		\begin{equation}\label{eq:Re_Sqp}
			\Re(\bar{S}_{qp}[\omega_0]) =  
			\frac{Q(\alpha_{q0},\alpha_{q0}) - 
				Q(\alpha_{q0}, 
				-\alpha_{q0})}{4|\alpha_{q0}|^2},
		\end{equation}
		\begin{equation}\label{eq:Im_Sqp}
			\Im(\bar{S}_{qp}[\omega_0]) =  
			\frac{Q(\alpha_{q0},-\ii\alpha_{q0}) - 
				Q(\alpha_{q0}, 
				\ii\alpha_{q0})}{4|\alpha_{q0}|^2}.
		\end{equation}
	\end{subequations}
	Note that in practice, it is not necessary to access the photocurrent
	spectrum at exactly $\omega = 0$; the above 
	expressions remain well-approximated over a bandwidth of frequencies for 
	which 
	$\boldsymbol{\bar{S}}[\omega]$ does not vary significantly. That is, if the 
	spectral feature of interest in $\boldsymbol{\bar{S}}$ is in a bandwidth  
	$\Delta\omega$ 
	around $\omega_0$, then
	\begin{equation}
		\bar{S}_{II}[\omega] \underset{\omega \ll \Delta\omega}{\approx} 
		\boldsymbol{\alpha_0}^\dagger 
		\SDMO{\omega_0}\boldsymbol{\alpha_0}
	\end{equation}
	confirming that synodyne detection is feasible.
	
	\subsection{Multimode symplectodyne detection}
	
	The discussion above can be naturally generalized to $n$ signals described 
	by their 
	quadratures $\boldsymbol{\hat{x}}(t)= [\hat{q}_1(t), \dots, \hat{q}_n(t), 
	\hat{p}_1(t), \dots, \hat{p}_n(t)]^T$, since 
	\cref{eq:symplectodyne_photocurrent,eq:symplectodyne_general_spectrum} for 
	the 
	photocurrent and its spectrum stay valid for a n-mode local oscillator 
	$\boldsymbol{\alpha}(t) = [\alpha_{q_1}(t), \dots, \alpha_{q_n}(t), \dots, 
	\alpha_{p_1}(t), \dots, \alpha_{p_n}(t)]^T \in\mathbb{R}^{2n}$. 
	The general photocurrent $\hat{I} = 
	\boldsymbol{\alpha}(t)^\dagger 
	\boldsymbol{\hat{x}}(t)$ can be realized (even without a multimode 
	combiner) by 
	performing synodyne detection on 
	each of the $n$ modes separately, yielding the single-mode photocurrents 
	$\hat{I}_m(t) = 
	\alpha_{q_m}(t) \hat{q}_m(t) + \alpha_{p_m}(t) \hat{p}_m(t)$, which are 
	themselves
	effectively classical stochastic processes, and therefore capable of being 
	recorded and
	combined electronically to obtain the desired sum photocurrent 
	$\hat{I}(t)$. 
	This scheme is shown in figure \cref{fig:symplectodyne}(b).

	\section{Applications}\label{sec:applications}
	
	To illustrate the physical insight of the conjugate symplectic group, we
	highlight various examples of physical systems on which the previous
	results apply. 
	
	\subsection{Single and two-mode synthesis} \label{sec:simple_synthesis}
	
	In case of a single mode, the conjugate symplectic group reduces to the 
	product of the
	circle group with the real symplectic group: $\SpH(2) = U(1)\times
	\text{Sp}(2n, \mathbb{R})$ as can be explicitly seen in the optical 
	decomposition theorem [\cref{thm:opt_decomp}]. This form represents the 
	frequency-dependent 
	rotation of squeezing, 
	as implemented in gravitational-wave detectors \cite{aligo_FDS,LIGO_SQL}. 
	The observable statistical moments are 
	contained in the single mode SDM, which is real-valued. Therefore, for a 
	single mode,
	the conjugate symplectic formulation is essentially equivalent to a real 
	symplectic treatment for every frequency. This is no longer true for two or 
	more modes.

	Indeed, for $n=2$, each matrix factor that appears in
	\cref{thm:opt_decomp} can be explicitly
	identified with a two-input/two-output optical
	component. The salient element is the frequency-dependent beam-splitter, 
	$\text{diag}(\boldsymbol{U}, \boldsymbol{U})$ with $\boldsymbol{U}\in 
	U(2)$, 
	which imprints complex rotations to the two-mode state. 
	(Such a beam-splitter can be realized using a critically-coupled
	resonator, see Supplementary Information 
	Section IX.)
	Thus the conjugate symplectic framework is
	essential and qualitatively distinct from the real-symplectic formalism, 
	since the cross-correlations between the quadratures 
	of a mode are now complex-valued in general.
	
	\subsection{Complex squeezing: its generation and diagnosis}
	\label{sec:dephasing_experiment}
	
	The inescapable physical manifestation of complex-valued cross-correlations 
	is the phenomenon
	of ``complex'' squeezing. This phenomenon, predicted in various quantum 
	systems 
	such as optomechanical devices \cite{Buchmann_2016,Ockeloen_Korppi_2018} 
	(including gravitational-wave
	detectors \cite{McCuller21}) and Kerr micro-resonators 
	\cite{Gouz23,Guidry:23,Dioum24}, 
	is still poorly understood within a systematic framework. 
	In particular, it is unknown what the sufficient requirement to produce 
	complex squeezing is, and
	how its presence can be diagnosed.
	Indeed, in gravitational-wave detectors, the precise failure to resolve 
	complex squeezing through 
	homodyne detection --- termed ``dephasing'' \cite{McCuller21,Kwee2014} --- 
	has been considered a 
	fundamental source of squeezing degradation.
	In the following, we show the sufficient condition to produce complex 
	squeezing, and how
	it can be uncovered by symplectodyne measurements.
	We define complex (or hidden) squeezing as follows. 
	
	\begin{definition}(Complex squeezing)
		Given a single-mode SDM $\boldsymbol{\bar{S}}[\omega]$, we say that the 
		underlying quantum state possesses 
		complex (or hidden) squeezing if the smallest eigenvalue of 
		$\boldsymbol{\bar{S}}$
		is lower than $1/2$, while the smallest eigenvalue of 
		$\Re(\boldsymbol{\bar{S}})$ is larger than $1/2$.
	\end{definition}
	
	Since a homodyne detector is only sensitive to $\Re(\boldsymbol{\bar{S}})$, 
	homodyne detectors
	are blind to complex squeezing; however, they can be resolved as sub-vacuum 
	fluctuations in a symplectodyne
	detector.
	
\begin{figure}
		\centering
		\includegraphics[width=\columnwidth]{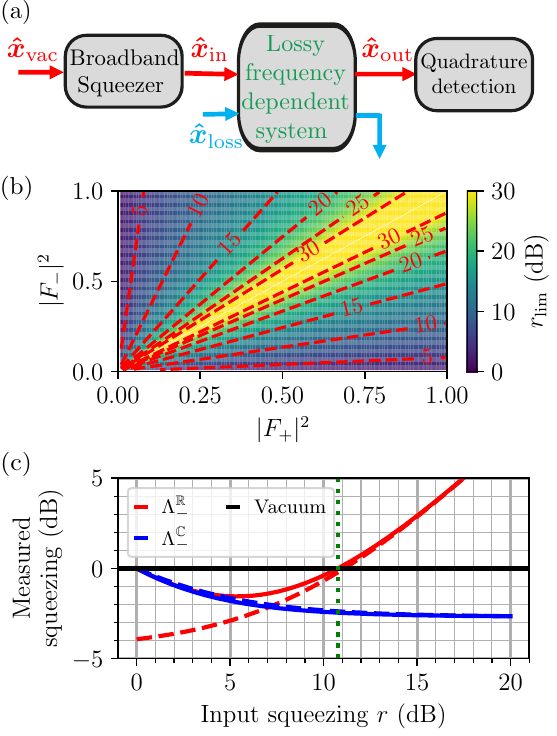}
		\caption{(a) General scheme for production of complex squeezing, and
			its detection.
			(b) Disappearance of squeezing in a homodyne detector
			due to frequency-dependent loss. The plot shows the limiting value 
			of
			input squeezing ($r_\text{lim}$ in \cref{eq:rlim}) in dB units 
			at which a homodyne detector fails to diagnose the presence of 
			squeezing due to
			frequency-dependent losses $F_+^2$ and $F_-^2$. 
			(c) Comparison of the optimal quadrature measurement 
			for a lossy 
			system described by \cref{eq:lossy_SDM_out}, using homodyne 
			($\Lambda_-^\mathbb{R}$, red) or synodyne detection 
			($\Lambda_-^\mathbb{C}$, blue). The sideband efficiencies are taken 
			to 
			be $F_+^2 = 0.99$ and $F_-^2 = 0.3$. The dashed lines are the 
			asymptotic
			expressions in \cref{eq:Lambda_lim}. The green 
			vertical line shows $r_\text{lim}$, above which the measured 
			squeezing becomes hidden.}
		\label{fig:complexsqzcavity}
	\end{figure}

	In order to understand the origin of complex squeezing, 
	consider a frequency-independent squeezed single mode, described by its SDM 
	$\boldsymbol{\bar{S}}_\text{in} = \text{diag}(e^{r}/2, e^{-r}/2)$, passing 
	through a lossy 
	passive system, as shown in \cref{fig:complexsqzcavity}(a). The 
	input-output relation is classically described by the relation 
	$a_\text{out}[\omega] = F[\omega]\, a_\text{in}[\omega]$ where 
	$|F[\omega]|\leq 
	1$. 
	Using the approach described in \ref{subsec:LTIdilated}, the quantum 
	description of such a system requires a single additional noise mode. The 
	output SDM is entirely determined by $F[\omega]$ and reads
	\begin{widetext}
		\begin{equation}\label{eq:lossy_SDM_out}
			\boldsymbol{\bar{S}}_\text{out}[\omega] =\frac{1}{2}\begin{bmatrix}
				1+(F_{+}^2 + F_-^2) \sinh^2r + F_- F_+ \cos(2\theta_D) 
				\sinh(2r) & F_+F_- \sin(2\theta_D) \sinh(2r) + \ii (F_+^2 - 
				F_-^2) \sinh^2r\\
				F_+F_- \sin(2\theta_D) \sinh(2r) - \ii (F_+^2 - 
				F_-^2) \sinh^2r & 1+(F_{+}^2 + F_-^2) \sinh^2r - F_- F_+ 
				\cos(2\theta_D) 
				\sinh(2r)
			\end{bmatrix} 
		\end{equation}
	\end{widetext}
	where $\theta_D[\omega] = \arg(F[\omega]F[-\omega])/2$, and we used the 
	shorthand notation $F_\pm \eqdef 
	|F[\pm\omega]|$. 
	The complex cross-correlations, described by the off-diagonal term $\pm \ii 
	(F_+^2 - F_-^2) \sinh^2r$, 
	become apparent when the losses are asymmetric in frequency
	($F_+\neq F_-$) and increase with the level of input squeezing $r$. 
	
	A physical system that presents such asymmetric-in-frequency loss for input 
	frequency-independent
	squeezing is a detuned critically-coupled optical cavity. The vacuum mode 
	coupled through the end 
	mirror 
	of the cavity combined with the resonance condition represent 
	the frequency-dependent loss, while the detuning induces the necessary 
	sideband 
	imbalance $F_+\neq F_-$ for complex squeezing. The magnitude of the cavity 
	response is
	\begin{align}
		|F[\omega]|^2 = \frac{4R \sin^2(\phi[\omega])}{(1-R)^2 + 4R 
			\sin^2(\phi[\omega])}
	\end{align}
	\noindent where $R$ is the power reflectivity of either mirror of the 
	cavity, 
	$\phi[\omega] = \omega L_0/c + \phi_0$ is the single-trip phase, with $L_0$ 
	the 
	resonant length of the cavity and $\phi_0\in[0, 2\pi]$ the detuning of the 
	cavity from the carrier resonance, which can be arbitrarily tuned by the 
	user. 
	Choosing $R$ and $\phi_0$ allows to realize any set of values for the 
	sideband 
	efficiencies $(F_+^2, F_-^2)\in[0, 1]^2$.
	
	In order to show that $\boldsymbol{\bar{S}}_\text{out}[\omega]$ displays 
	complex squeezing, we compare the lowest eigenvalues of 
	$\boldsymbol{\bar{S}}_\text{out}[\omega]$ and 
	$\Re(\boldsymbol{\bar{S}}_\text{out}[\omega])$, namely 
	$\Lambda_-^{\mathbb{C}}$ 
	and $\Lambda_-^{\mathbb{R}}$. 
	As per the discussion in \cref{sec:realization_SDM}, a symplectodyne 
	(respectively, homodyne) detector 
	is sensitive to the former (latter).
	These eigenvalues are shown in \cref{fig:complexsqzcavity}(c): as the input 
	squeezing level
	increases, the homodyne detector is oblivious to the presence of squeezing, 
	whereas a symplectodyne detector
	resolves it below the vacuum level. Thus the state is complex squeezed.
	Indeed, it can be shown (see Supplementary Information 
	Section X) that 
	in the limit of high squeezing, 
	\begin{equation}\label{eq:Lambda_lim}
		\begin{split}
			\Lambda_-^{\mathbb{C}} &\xrightarrow[r\to\infty]{} \frac{1}{2} 
			\left[ 
			(1-\eta^\mathbb{C}) + 
			\eta^\mathbb{C} e^{-2r}\right] < \frac{1}{2}\\
			\Lambda_-^{\mathbb{R}} & \xrightarrow[r\to\infty]{}\frac{1}{8} (F_+ 
			- 
			F_-)^2 e^{2r} \gg \frac{1}{2}
		\end{split}
	\end{equation}
	\noindent where $\eta^\mathbb{C} = 
	{2}/({F_-^{-2} + F_+^{-2}})$ is the \textit{harmonic mean} of the single 
	sideband efficiencies at $\pm\omega$.
	
	The threshold of input squeezing $r_\text{lim}$ for which 
	squeezing becomes hidden, obtained by solving for 
	$\Lambda_-^{\mathbb{R}}(r_\text{lim})=1/2$, yields the simple form
	\begin{equation}\label{eq:rlim}
		r_\text{lim} = \text{arctanh}\left(\frac{2F_+F_-}{F_+^2+F_-^2}\right) 
	\end{equation}
	For input squeezing level $r > r_\t{lim}$, the squeezing properties present 
	in 
	$\boldsymbol{\bar{S}}_\text{out}$ are
	completely hidden in the complex correlations, and homodyne detection fails 
	at 
	identifying this quantum resource, see \cref{fig:complexsqzcavity}(b). 
	Worse, it identifies the state in question 
	as having increasingly large classical fluctuations (scaling as $e^{2r}$) 
	even 
	in the optimal quadrature, whereas a 
	symplectodyne detector accurately diagnoses its quantum nature.
	
	Importantly, the apparent loss of squeezing due to frequency-dependent loss
	(observed in gravitational-wave detectors \cite{McCuller21}) is not a 
	fundamental limit 
	but rather a penalty due to the choice of sub-optimal detection schemes. 
	Similarly, other quantum information schemes 
	which require high levels of optical squeezing 
	\cite{braunstein00,menicucci2014,bourassa21} can have apparent 
	degradation of quantum resources due to dispersive lossy components, unless 
	a proper 
	detection scheme is employed.
	
	The scheme sketched above applies to frequency-dependent \emph{loss}; but 
	similar phenomenology
	also applies to frequency-dependent \emph{gain}, such as in
	phase-mixing amplifiers \cite{Ockeloen-Korppi_2017}, though their 
	minimal description requires up to two additional noise modes. 
	
	\subsection{Tight quantum noise limits for linear feedback oscillators: The 
	Schawlow-Townes limit}

	\begin{figure}[t!]
		\centering
		\includegraphics[width=0.9\columnwidth]{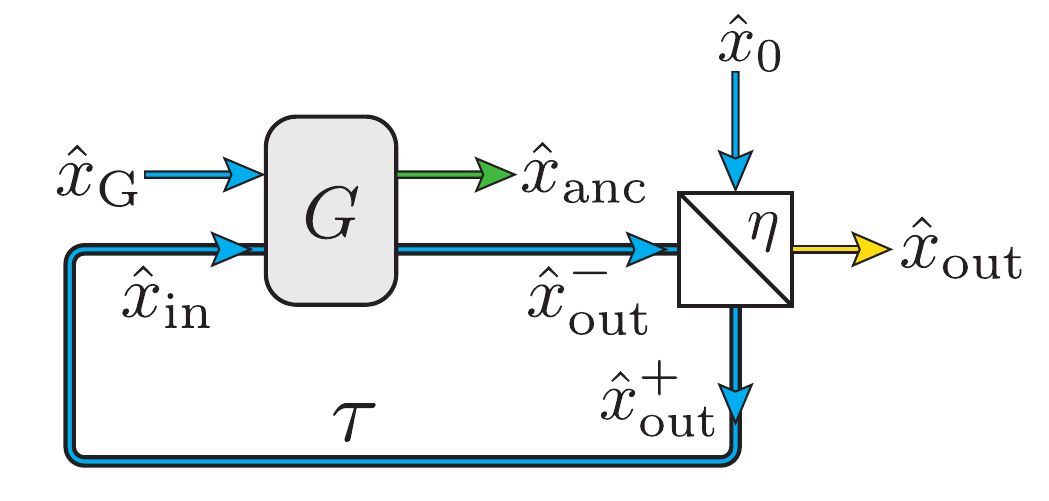}
		\caption{\label{fig:fbOscSchematic} A quantum feedback oscillator in 
		which the output of an amplifier, labeled $G$, is partially out-coupled 
		and partially fed back to its input by a beam splitter labeled $\eta$ 
		after a time delay $\tau$. For most physical feedback oscillators, such 
		as lasers, the quantum noise inputs $\hat{x}_0$ and $\hat{x}_\text{G}$ 
		to the beam splitter and amplifier are inaccessible, the amplifier's 
		output, $\hat{x}_\text{anc}$ is inaccessible, and the field out-coupled 
		by the beam splitter, $\hat{x}_\text{out}$ gives the system's 
		accessible output.}
	\end{figure}
	
	The quantum noise performance of a laser, described by the Schawlow-Townes 
	formula, 
	is a fundamental problem in quantum optics. It has been shown 
	\cite{Loughlin23} that the 
	Schawlow-Townes formula is a general feature of all feedback oscillators, 
	and that it can be evaded by 
	quantum engineering the amplifier and out-coupler in the feedback loop of 
	an oscillator. But the question
	of whether it is possible to suppress output fluctuations all the way to 
	the minimum level imposed 
	by the Heisenberg uncertainty bound [\cref{eq:tightUncertainty}] remained 
	open.
	
	In order to settle this question in the affirmative, we revisit the simple 
	model of a
	feedback oscillator as a two-input, two-output device (see 
	\cref{fig:fbOscSchematic})
	where the device's outputs carry quantum fluctuations on top of a mean
	macroscopic signal as depicted in \cref{fig:fbOscSchematic}.
	The device's observed output $\hat{x}_\text{out}$ and unobserved output 
	$\hat{x}_\text{anc}$, are 
	related to the noise inputs to the beam splitter, $\hat{x}_0$, and 
	amplifier, $\hat{x}_\text{G}$, by \cite{Loughlin23},
	$[\hat{q}_\text{out}, \hat{q}_\text{anc},
	\hat{p}_\text{out}, \hat{p}_\text{anc}]^\T = \boldsymbol{M} 
	[\hat{q}_0, \hat{q}_\text{G},
	\hat{p}_0, \hat{p}_\text{G}]^\T$
	where the transfer matrix
	\begin{equation}\label{eq:fullFbOscTf}
			\boldsymbol{M}[\omega] = \begin{bmatrix}
				H_0[\omega] & H_\text{G}[\omega] & 0 & 0 \\
				e^{\ii\omega\tau}H_\text{G}[\omega] & H_\text{A}[\omega] & 0 &
				0 \\
				0 & 0 & H_0[\omega] & -H_\text{G}[\omega]  \\
				0 & 0 & -e^{\ii\omega\tau} H_\text{G}[\omega] &
				H_\text{A}[\omega]
			\end{bmatrix}
	\end{equation}
	can be expressed in terms of the transfer functions
	\begin{equation}\label{amp:H0HG}
		\begin{split}
			H_0[\omega] &=
			\frac{\sqrt{\eta}+e^{i \omega \tau}/\sqrt{\eta}}{1+e^{\ii \omega
					\tau}} \\
			H_\text{G}[\omega] &=
			\frac{1/\sqrt{\eta} - \sqrt{\eta}}{1+e^{\ii \omega \tau}} \\
			H_\text{A}[\omega] &=
			\frac{1/\sqrt{\eta} + \sqrt{\eta} e^{\ii \omega \tau}}{1+e^{\ii
					\omega
					\tau}},
		\end{split}
	\end{equation}
	and $\eta$ is the beam-splitter's power reflectivity, $G = 1/\sqrt{\eta}$ 
	is the amplifier's amplitude gain, and $\tau$ is the round trip time delay.
	It can be verified that $\boldsymbol{M}[\omega]
	\boldsymbol{J} \boldsymbol{M}[\omega]^\dagger = \boldsymbol{J}$. 
	
	The conjugate symplectic character of $\boldsymbol{M}$ immediately implies 
	that $\boldsymbol{M}^{-1}$
	also exists as a conjugate symplectic matrix. The corresponding inverse 
	system can be physically implemented
	if $\boldsymbol{M}^{-1}$ is causal. This is indeed the case since
	\begin{equation}
		\boldsymbol{M}[\omega]^{-1} =
		\begin{bmatrix}
			H_\text{A}[\omega] & -H_\text{G}[\omega] & 0 & 0 \\
			
			-e^{\ii\omega\tau}H_\text{G}[\omega] & H_0[\omega] & 0 & 0 \\
			0 & 0 & H_\text{A}[\omega] & H_\text{G}[\omega]  \\
			0 & 0 & e^{\ii\omega\tau} H_\text{G}[\omega] & H_0[\omega]
		\end{bmatrix}
	\end{equation}
	involves the same transfer functions as those in $\boldsymbol{M}[\omega]$, 
	each of which is causal. 
	Then, the protocol to saturate the Heisenberg bound for the feedback 
	oscillator is to
	act with the causal conjugate symplectic matrix 
	$\boldsymbol{M}[\omega]^{-1}$ on the vacuum inputs
	of the amplifier and beam-splitter. The feedback oscillator's output mode 
	will then
	have the quantum fluctuations of a vacuum state.
	
	On the other hand, if the input noise modes $\hat{x}_\text{in}$ and
	$\hat{x}_\text{G}$ are uncorrelated, we can apply the uncertainty bound of 
	\cref{eq:generalizedUncertaintyNoiseOnly}
	with $\{\boldsymbol{N}_{ij}\} = \{\boldsymbol{N}_{\text{out},0}, 
	\boldsymbol{N}_{\text{out},\text{G}}\} $ where these matrices
	can be read off from \cref{eq:fullFbOscTf}.
	We find
	\begin{equation}
		\begin{split}
			\bar{S}_{qq}^\text{out}[\omega] \bar{S}_{pp}^\text{out}[\omega] 
			\geq& 
			\frac{1}{4} \Big(|\det(\boldsymbol{N}_{\text{out},0})| + 
			|\det(\boldsymbol{N}_{\text{out},\text{G}})| \Big)^2 \\
			\geq& \left(|H_\text{G}[\omega]|^2 + \frac{1}{2} \right)^2,
		\end{split}
	\end{equation}
	where we have used $|H_0[\omega]|^2 = |H_\text{G}[\omega]|^2+1$ to simplify 
	the 
	final expression.
	This bound generalizes the Schawlow-Townes limit in the spectral domain at 
	all frequencies, and exactly recovers it around the carrier frequency
	when quantum fluctuations are distributed equally between the output 
	quadratures.
	Notably, since the matrices 
	$\boldsymbol{N}_{\text{out},0}\boldsymbol{N}_{\text{out},0}^\dagger$ and 
	$\boldsymbol{N}_{\text{out},\text{G}}\boldsymbol{N}_{\text{out},\text{G}}^\dagger$
	 are linearly dependent, the Minkowski inequality becomes equality and this 
	lower bound is achievable. Indeed, using \cref{eq:fullFbOscTf} and assuming 
	the noise modes are in vacuum states, we see that this uncertainty bound 
	saturated,
	so this lower bound is tight. We note that this tight lower bound is 
	readily proved using the conjugate-symplectic formalism developed in this 
	paper, yet evaded the previous analysis of ref. \cite{Loughlin23}, which 
	proved a slightly weaker bound.
	
	\section{Conclusion}
	
	We have developed the first complete theory of quantum linear 
	time-translation-invariant systems, effectively providing a means to arrive 
	at a 
	consistent quantum description of a classical LTI system.
	In particular, this quantization process proceeds by promoting a classical 
	LTI system
	to a quantum LTI system with additional quantum noise; we derived the 
	minimal level of
	added quantum noise required to enable this transformation.
	The result is in general an open quantum system.
	We established a procedure to dilate it into a closed quantum LTI system 
	which evolves 
	unitarily and is described by an element of the conjugate symplectic 
	group $\SpH(2n)$.
	
	The conjugate symplectic group is a Lie group, which subsumes the 
	oft-studied
	real-symplectic group in the realm of Gaussian quantum information theory.
	We elucidated the structure of the conjugate symplectic group, in 
	particular its
	one-parameter unitary representation, which gives the Hamiltonian 
	describing these systems.
	We proved a decomposition theorem that enables synthesis of an arbitrary 
	quantum LTI system 
	in terms of readily constructed optical elements. This synthesis procedure 
	extends and completes
	the theory of network synthesis of classical LTI systems.
	
	The group structure of quantum LTI systems also naturally gives rise to 
	spectral uncertainty bounds for open and closed systems, which are stricter 
	than the ones usually derived for frequency independent systems.
	
	We established a canonical invariant representation of the spectral density 
	matrices of the outputs
	of quantum LTI systems. This generalizes Williamson's theorem, and 
	identifies a broader 
	class of symplectic invariants that involves a novel form of complex 
	quantum correlation.
	These complex correlations manifest as complex squeezing, a quantum 
	resource that has 
	evaded systematic analysis thus far. Complex squeezing is 
	undetectable with standard homodyne or heterodyne detectors and require 
	more 
	general symplectodyne detection to be fully revealed, whose theory we 
	established.
	
	With a full theory of quantum LTI systems in hand, we explored several of 
	its 
	immediate implications. 
	In laser theory, this framework immediately results in tight uncertainty 
	bounds 
	for laser stability that generalize the Schawlow-Townes limit, and shows 
	how 
	lasers may be optimally engineered to evade 
	classical bounds on their performance. Similarly, our theory explains the 
	precise origin and mechanism of ``dephasing'' loss of squeezing due to 
	frequency-dependent lossy systems such as 
	gravitational wave detectors, and how increased squeezing levels may be 
	recovered with optimal detection schemes.
	
	All of this is done without reference to any internal model of the system 
	and therefore applies directly
	to the measured response of a quantum system without any need for system 
	identification.
	That should endow the user of this framework with the power to analyze, 
	synthesize, measure, and generally 
	make sense of complex quantum systems as found in their natural laboratory 
	habitat.

	\begin{acknowledgments}
		J.D.  gratefully acknowledges the support of the EU Horizon 2020 
		Research 
		and Innovation Program under the Marie Sklodowska-Curie
		Grant Agreement No. 101003460 (PROBES).
		H. A. L. gratefully acknowledges the support of the National Science 
		Foundation through the LIGO operations cooperative agreement 
		PHY18671764464.
	\end{acknowledgments}

	
\appendix

\section{Notations and conventions}\label{app:notations}

This appendix details the notations and conventions used throughout this 
article. 

The identity matrix in $\mathbb{R}^n$ is written 
$\boldsymbol{1}_n$. For a matrix $\boldsymbol{M}$, its transpose is
denoted $\boldsymbol{M}^T$, element-wise complex conjugate is $\boldsymbol{M}^*$, and
conjugate transpose is $\boldsymbol{M}^\dagger = (\boldsymbol{M}^*)^T$. 
For a single operator $\hat{O}$ living in a Hilbert space, its
adjoint is denoted $\hat{O}^\dagger$. That is, we distinguish between the 
different spaces on which quantum operators and matrices act. 
For matrices, the abbreviation ``HP(S)D'' stands for hermitian positive (semi) 
definite.

\subsection{Operators in time and frequency domain 
	\label{subsection:operators_FT}}

\begin{definition}[Fourier transform of an operator]
	For any operator in time domain $\hat{A}(t)$, we define its Fourier 
	transform $\hat{A}[\omega]$ as
	\begin{equation}
		\hat{A}[\omega] = \int_{-\infty}^{+\infty} \hat{A}(t) e^{\ii\omega t} 
		\dd t 
	\end{equation}
	\noindent or equivalently
	\begin{equation}
		\hat{A}(t) = \int_{-\infty}^{+\infty} \hat{A}[\omega] e^{-\ii\omega 
			t} \frac{\dd \omega}{2\pi}.
	\end{equation}
\end{definition}

We will denote the adjoint of $\hat{A}(t)$ by $\hat{A}^\dagger(t)$ or 
$\hat{A}(t)^\dagger$;
however, we distinguish between $\hat{A}^\dagger [\omega]$ which is the Fourier 
transform of
$A^\dagger(t)$, and $\hat{A}[\omega]^\dagger$ which is the adjoint of the 
Fourier transform 
of $\hat{A}(t)$. These are related as $\hat{A}^\dagger[\omega] = 
\hat{A}[-\omega]^\dagger$. 
For a self-adjoint operator $\hat{X}(t)$, i.e. $\hat{X}(t)=\hat{X}^\dagger(t)$, 
we further
have $\hat{X}^\dagger[\omega] = \hat{X}[-\omega]^\dagger = \hat{X}[\omega]$. 
These remarks mirror the properties for scalar functions and their Fourier 
transforms: for any function $f : t\mapsto f(t)\in\mathbb{C}$, we denote the 
complex conjugate of $f(t)$ by $f(t)^*$ or $f^*(t)$ interchangeably; in 
frequency domain, however, we 
have for any $\omega \in\mathbb{R}$ that $f[-\omega]^*  = f^*[\omega]$. If 
$f(t)\in\mathbb{R}$ for any $t\in\mathbb{R}$, then $f^*[\omega] =f[-\omega]^* = 
f[\omega]$.  

The convolution of two scalar functions $f$ and $g$ is
\begin{equation}
	\begin{split}
		(f\star g)(t) &\eqdef \int_{-\infty}^{+\infty} f(\tau) g(t-\tau)\dd\tau 
		\\
		&= \int_{-\infty}^{+\infty} f[\omega]g[\omega]e^{-\ii\omega t} 
		\frac{\dd\omega}{2\pi}.
	\end{split}
\end{equation}

Likewise, if $h(t) \eqdef f^*(t)g(t)$ (beware that the complex conjugate is 
on the time-domain function $f$), then its Fourier transform is
\begin{equation}\label{eq:convolution_conjugate}
	h[\omega] = \int_{-\infty}^{+\infty}f[\Omega]^* g[\omega + \Omega] 
	\frac{\dd\Omega}{2\pi}.
\end{equation}


We also recall the Parseval-Plancherel theorem, which can be extended to 
well-behaved time-domain operators.
\begin{theorem}\label{thm:Parseval_Plancherel}
	Let $\hat{A}(t)$ a time-domain operator and $f$ a well-behaved 
	function. Then,
	\begin{equation}
		\int_{-\infty}^{+\infty}f^*(t) \hat{A}(t) \dd t = 
		\int_{-\infty}^{+\infty}f[\omega]^* \hat{A}[\omega] \frac{\dd 
			\omega}{2\pi}.
	\end{equation}
\end{theorem}

\subsection{Correlation and spectral density matrices}\label{app:SDM}

The cross-correlation matrix $\boldsymbol{\bar{S}}(t)$ for the quadratures 
of an $n$-mode system is defined as   
\begin{equation}
	\bar{S}_{jk}(t) = \frac{1}{2} 
	\left\langle\left\{(\boldsymbol{\hat{x}}_j(t) - \langle 
	\boldsymbol{\hat{x}}_j(t)\rangle), (\boldsymbol{\hat{x}}_k(0) - \langle 
	\boldsymbol{\hat{x}}_k(0)\rangle)\right\}  \right\rangle.
\end{equation}
Assuming that $\langle \boldsymbol{\hat{x}}_j(t)\rangle = 0$, which we will assume to be true unless specified otherwise,
$\bar{S}_{jk}(t) = \frac{1}{2} 
\left\langle\left\{\boldsymbol{\hat{x}}_j(t), 
\boldsymbol{\hat{x}}_k(0)\right\}
\right\rangle$. For the vacuum state
\begin{equation}\label{eq:vacuum_cross_correlation}
	\boldsymbol{\bar{S}}_\text{vac}(t) = \frac{\boldsymbol{1}}{2} \delta(t).
\end{equation}


%
%
%

The \textit{spectral density matrix} (SDM), 
$\boldsymbol{\bar{S}}[\omega]$, is the element-wise Fourier transform of the 
cross-correlation matrix in the weak-stationary regime: 
$\boldsymbol{\bar{S}}[\omega] = 
\int_{-\infty}^{+\infty} 
	\boldsymbol{\bar{S}}(t) 
	e^{\ii\omega 
		t} \dd t$.
Equivalently,
\begin{equation}\label{eq:def_sdm}
	\qquad \bar{S}_{ij}[\omega] \times 2\pi\delta[\omega-\omega'] = \frac{1}{2} 
	\left\langle\left\{\hat{x}_i[\omega], 
	\hat{x}_j[\omega]^\dagger\right\}  \right\rangle.
\end{equation}
The following theorem illustrates important properties of the SDM.

\begin{theorem}[Properties of the SDM] \label{thm:SDM_SDP}
	For any frequency $\omega\in \mathbb{R}$, $\SDMO{\omega}$ is a complex 
	valued, hermitian positive semi-definite matrix and $\SDMO{\omega}^* = 
	\SDMO{-\omega}$ (where $(\cdot)^*$ is the element-wise complex conjugate).
\end{theorem}

\begin{proof}
	The hermitian nature of the SDM is due to the hermitian symmetry of the dot 
	product. The equality $\SDMO{\omega}^* = \SDMO{-\omega}$ can be checked 
	explicitly.  
	Let us show that the SDM is positive semi-definite. 
	Let  $\boldsymbol{\alpha} = 
	[\alpha_1, \dots \alpha_{2N}]^T \in \mathbb{C}^{2N}$ and $\hat{A}[\omega] = 
	\alpha_j^* \hat{x}_j[\omega]$. Then
	\begin{equation}
		\boldsymbol{\alpha}^\dagger \bar{\boldsymbol{S}}\boldsymbol{\alpha} = 
		\frac{1}{4\pi \delta[0]} 
		\left\langle\left\{\alpha^*_j\hat{x}_j[\omega], 
		\alpha_k\hat{x}_k[\omega]^\dagger\right\}
		\right\rangle = \bar{S}_{AA}[\omega] \geq 0.
	\end{equation}
\end{proof}

So far the results are valid without invoking the commutation relations of 
\cref{eq:ccrAndJdef} on the quadrature vector. 
Taking this into account 
yields the uncertainty principle $\SDMO{\omega} + \ii \boldsymbol{J}/2 \geq 0$ 
of \cref{thm:up_nmode}. One can also show that 
$\SDMO{\omega}$ is hermitian positive definite at all frequencies, as detailed 
in SI Section V.

The utility of the SDM is due to the fact that
under an LTI transformation of weak-stationary noises of the form 
\cref{eq:transfer_function_matrix_def}, the associated SDM transforms as
\begin{equation}
	\boldsymbol{\bar{S}}_\text{out}[\omega] = \boldsymbol{M}[\omega] 
	\boldsymbol{\bar{S}}_\text{in}[\omega] \boldsymbol{M}[\omega]^\dagger.
\end{equation}

The SDM can also be expressed in terms of the ladder operators
$\boldsymbol{\hat{\mathfrak{a}}}[\omega]$ as
\begin{equation}\label{eq:def_SDM_a} 
	\bar{S}^\mathfrak{a}_{jk}[\omega] \times 2\pi \delta[0] \eqdef \frac{1}{2} 
	\left\langle \left\{ \hat{\mathfrak{a}}_j[-\omega]^\dagger, 
	\hat{\mathfrak{a}}_k[-\omega]\right\}\right\rangle.
\end{equation}
While $\boldsymbol{\bar{S}}^\mathfrak{a}[\omega]$ is hermitian,
the non-hermiticity of the ladder operators imply that,
$\boldsymbol{\bar{S}}^\mathfrak{a}[-\omega]^*\neq 
\boldsymbol{\bar{S}}^\mathfrak{a}[\omega]$, contrary to $\SDM$ (see 
\cref{thm:SDM_SDP}). They are nevertheless related through $		
\SDMO{\omega} = 
\boldsymbol{P}^* 
\boldsymbol{\bar{S}}^\mathfrak{a}[\omega] \boldsymbol{P}^T = 
\boldsymbol{P} \boldsymbol{\bar{S}}^\mathfrak{a}[-\omega]^* 
\boldsymbol{P}^\dagger$, which can be shown by direct computation.
Using the commutation relations gives the useful expression
\begin{equation}\label{eq:simplified_SDM_a}
	\bar{S}^\mathfrak{a}_{jk}[\omega] = \frac{1}{2\pi\delta[0]} 
	\left\langle \hat{\mathfrak{a}}_j[-\omega]^\dagger 
	\hat{\mathfrak{a}}_k[-\omega]\right\rangle + 
	\frac{(\boldsymbol{I}_n)_{jk}}{2}.
\end{equation}

\section{Input-output description of a classical LTI system}\label{app:ltiClassical}

We take the input-output response of a classical system to be defined by
a sequence of maps $\{\mathbfcal{G}^t\}$ whose action $
	\boldsymbol{x}^\t{out}(t) = \mathbfcal{G}^t[\boldsymbol{x}^\t{in}]$
is to map the input signal $\boldsymbol{x}^\t{in}=\{x^\t{in}_i\}$ (at all 
times) to the output
signal value $\boldsymbol{x}^\t{out}(t)$ at the instant $t$.
Thus the system is equivalent to the set $\{\mathbfcal{G}^t\}$ of multi-input 
multi-output maps.

\begin{lemma}\label{thm:MimoToSiso}
	A multi-input, multi-output, linear map $\mathbfcal{G}^t = 
	\{\mathcal{G}^t_i\}$, 
	which maps a collection of inputs to a collection of outputs can be 
	decomposed as a 
	collection of linear maps from a single input to a single output.
\end{lemma}
\begin{proof}
	For a system with $m$ outputs, the input-output relation 
	$\boldsymbol{x}^\text{out}(t) = \mathbfcal{G}^t[\boldsymbol{x}^\text{in}]$ 
	represents a collection of
	$m$ equations, the $i^\text{th}$ of which is $
		x^\text{out}_i(t) = \mathcal{G}_i^t[\boldsymbol{x}^\text{in}],
	$
	where $\mathcal{G}_i^t$ is the map from the collection of inputs to the 
	$i^\text{th}$ output. 
	Let $\boldsymbol{e}_j $ be the vector with $1$ in the $j^\t{th}$ position 
	and zero elsewhere;
	so, $\boldsymbol{x}^\text{in} = \sum_j x^\text{in}_j \boldsymbol{e}_j$. 
	Then by linearity
	\begin{equation}\label{eq:splitLinearInputs}
		\begin{split}
			x^\text{out}_i(t) &= \mathcal{G}^t_i\left[\sum_j x^\text{in}_j 
			\boldsymbol{e}_j\right] \\
			&= \sum_j \mathcal{G}^t_i\left[x^\text{in}_j \boldsymbol{e}_j\right]
			\defeq \sum_j \mathcal{G}^t_{ij}\left[x_j^\text{in} \right].
		\end{split}
	\end{equation}
	Thus the $i^\t{th}$ output is a linear superposition of each of the $j$ 
	inputs
	via the map $\mathcal{G}^t_{ij}$.
	The fact that $\mathcal{G}^t_{ij}$ is linear follows by setting all inputs 
	to zero except the $j^\t{th}$
	in \cref{eq:classicalLinearity}. 
\end{proof}

Thus a linear system is represented by a set of single-input single-output 
linear maps $\{\mathcal{G}^t_{ij}\}$. 
If we assume that the input signals form a Hilbert space, then the  
Riesz representation theorem shows
that each of the maps can be uniquely represented by a vector in the dual of 
the Hilbert space.

\begin{theorem}[Riesz \cite{Yosida80}]\label{thm:sisoLti}
	In a Hilbert space $\mathscr{H}$ equipped with the inner product 
	$\langle\cdot,\cdot\rangle$,
	every linear map $\mathcal{G}$ on it
	is uniquely represented as $\mathcal{G}[x] 
	= \langle G, x \rangle$.
\end{theorem}
\begin{proof}
	Note first that by the linearity of $\mathcal{G}$:
	\begin{equation}
		\mathcal{G}\big[ \mathcal{G}[x]y - x \mathcal{G}[y] \big]
		= \mathcal{G}[x]\mathcal{G}[y] - \mathcal{G}[x]\mathcal{G}[y]
		= 0;
	\end{equation}
	so $\mathcal{G}[x]y - x \mathcal{G}[y] \in \t{ker} \mathcal{G}$, the kernel 
	space of $\mathcal{G}$. Now choose $y \in (\ker \mathcal{G})^\perp$ such that 
	$\langle y, y \rangle = 1$; then
	\begin{equation}
		\begin{split}
			\mathcal{G}[x] &= \mathcal{G}[x]\langle y, y
			\rangle 
			\\
			&= \langle y, \mathcal{G}[x] y \rangle \\
			&= \langle y, \mathcal{G}[x] y - x
			\mathcal{G}[y]  
			\rangle + \langle y, x \mathcal{G}[y] \rangle.
		\end{split}
	\end{equation}
	The first term is zero since $y$ and $\mathcal{G}[x] y - x\mathcal{G}[y]$ 
	lie in orthogonal spaces. Thus
	\begin{equation}\label{eq:elementLinearity}
		\begin{split}
			\mathcal{G}[x] &= \langle y, x \mathcal{G}[y] 
			\rangle 
			= \langle \mathcal{G}[y]^* y, x  \rangle \defeq 
			\langle 
			G, x \rangle.
		\end{split}
	\end{equation}
	For uniqueness, suppose $\mathcal{G}[x] = \langle G,x 
	\rangle 
	= \langle G^\prime, x \rangle$, then $\langle G - 
	G^\prime,x \rangle = 0$ $\forall x$, so $G - G^\prime =0$ 
	and $G$ is unique.
\end{proof}

Using \cref{thm:MimoToSiso}, we can extend \cref{thm:sisoLti} to a multi-input, 
multi-output maps.
To wit, using \cref{eq:elementLinearity,eq:splitLinearInputs}, we have
\begin{equation}\label{app:ltiInOutMatM}
	x^\text{out}_i(t) = \sum_j \mathcal{G}^t_{ij}\left[x_j^\text{in} \right] 
	= \sum_j \langle G_{ij}^t, x^\text{in}_j \rangle.
\end{equation}
Thus, the set of maps $\{\mathbfcal{G}^t\}$ is equivalent to
the set of matrices $\{\boldsymbol{G}^t\}$ for a linear system. 

For the rest of this section, whenever matrices and vectors are involved, we 
use the summation convention for indices.

For real-valued signals of finite energy, the natural Hilbert 
space is $\mathscr{H} = L^2(\mathbb{R})$, for which the inner product is 
$\langle y,x\rangle = \int y(t) x(t)\dd t$.
In this case \cref{app:ltiInOutMatM} becomes
\begin{equation}\label{app:ltiInOutInteg1}
	\begin{split}
		x_i^\text{out}(t) &= \mathcal{G}_{ij}^t [x_j^\t{in}] = \langle G^t_{ij}, x_j^\text{in} \rangle \\
		&= \int G_{ij}^t(t^\prime) x_j^\text{in}(t^\prime) \dd t^\prime 
		\defeq \int G_{ij}(t,t^\prime) x_j^\text{in}(t^\prime)\dd t^\prime.
	\end{split}
\end{equation}
That is, the set of linear maps $\{\mathbfcal{G}^t\}$ is equivalent to a set of 
matrices $\{\boldsymbol{G}(t,t')\}$ for each pair of instances $t,t'$.

Next we 
 show that a linear time-translation-invariant (LTI) system's matrices 
$\{\boldsymbol{G}(t,t')\}$ are only functions of the time difference $t-t'$, in which case 
the integral input-output relation in \cref{app:ltiInOutInteg1} becomes a 
convolution. Recall the definition of a LTI system in \cref{def:LTI}.

\begin{theorem}
	The input-output matrices $\{\boldsymbol{G}(t,t')\}$ of a LTI system assume 
	the form $\boldsymbol{G}(t,t^\prime) = \boldsymbol{G}(t-t^\prime)$, so the 
	input-output response [\cref{app:ltiInOutInteg1}] reduces to a convolution
	\begin{equation}\label{eq:ltiInOutTime}
		x_i^\t{out}(t) = \mathcal{G}_{ij}^t [x_j^\t{in}] 
		= \int G_{ij}(t - t^\prime) x_j^\text{in}(t^\prime)\dd t^\prime.
	\end{equation}
\end{theorem}
\begin{proof}
	By the definition of time-translation invariance:  
	$\mathcal{G}_{ij}^t[\mathcal{T}^s[x_j^\text{in}(t^\prime)]] 
	= \mathcal{T}^s[\mathcal{G}_{ij}^t[x_j^\text{in}(t^\prime)]]$, so
	\begin{equation}
		\int G_{ij}(t,t^\prime + s) x_j^\text{in}(t^\prime)\dd t^\prime = 
		\int G_{ij}(t-s,t^\prime) x_j^\text{in}(t^\prime) \dd t^\prime
	\end{equation}
	for all $x_j^\text{in}(t^\prime)$, 
	which implies $G_{ij}(t,t^\prime + s) = G_{ij}(t-s,t^\prime)$. Thus,
	$G_{ij}(t,s) = G_{ij}(t,0+s) = G_{ij}(t-s,0) \defeq G_{ij}(t-s)$.
\end{proof}

A LTI system's input-output relation is naturally written in the time 
domain as a convolution as in \cref{eq:ltiInOutTime}. 
Taking the Fourier transform of \cref{eq:ltiInOutTime}, we 
have
\begin{equation}
	\boldsymbol{x}^\text{out}[\omega] = \boldsymbol{G}[\omega] 
	\boldsymbol{x}^\text{in}[\omega].
\end{equation}
Thus any classical LTI system has outputs linearly related to its inputs in 
the frequency domain, where the relation between inputs and outputs is 
governed by the frequency-dependent transfer function matrix 
$\boldsymbol{G}[\omega]$.

\section{Algorithms for the optical decomposition}
\label{app:algorithms}

The algorithms below implement the optical decomposition at any arbitrary 
frequency. They are written in pseudo code but use the NumPy indexing and 
slicing conventions.
\begin{itemize}
	\item \cref{algocf:alg_symplectic_sort}, \textsc{SortEigenelements}, sorts 
	the eigenelements of an input 
	matrix in HPD$\cap\SpH(2n)$. We assume that the 
	function \textsc{Diagonalize}($\boldsymbol{M}$) returns an unordered 
	list $\boldsymbol{\lambda}$ of eigenvalues (with multiplicities) and 
	the corresponding unitary matrix $\boldsymbol{V}$ with eigenvectors as 
	columns. It sorts the eigenelements to be compatible with the ordering 
	$\boldsymbol{M} = \boldsymbol{V_0}\begin{bmatrix}
		\text{diag}(\boldsymbol{\ell}) & \boldsymbol{0}\\
		\boldsymbol{0} & \text{diag}(\boldsymbol{\ell})^{-1}\\
	\end{bmatrix} \boldsymbol{V_0}^\dagger$ where $\boldsymbol{V_0} 
	\in U(2n)$ and $\boldsymbol{\ell}\in\left(\mathbb{R}^*_+\right)^n$ 
	ordered 
	in increasing order as $1\leq
	\ell_1\leq \dots\leq \ell_n$ (see ref. \cite{elie_thesis} page 136).
	
	\item \cref{algocf:alg_CSD}, \textsc{CsdSp}, performs the conjugate 
	symplectic CSD of an 
	input matrix which is conjugate symplectic and unitary (see SI 
	Theorem 8).
	
	\item \cref{algocf:alg_symplectic_HPD}, \textsc{SymplecticSpectral}, 
	performs the spectral decomposition 
	in HPD$\cap\SpH(2n)$ (see SI 
	Theorem 7, and 
	ref. \cite{elie_thesis} page 136) 
	\item \cref{algocf:alg_bloch_messiah}, \textsc{SvdSp}, does the 
	Bloch-Messiah decomposition 
	in $\SpH(2n)$ 
	(see SI Theorem 7, ref. 
	\cite{elie_thesis} page 136, and ref. \cite{Gouzien20}). 
	$\textsc{Polar}(\boldsymbol{M})$ returns the unique right-polar 
	decomposition of an invertible matrix $\boldsymbol{M}$.
\end{itemize}

\begin{algorithm}[H]
	\caption{\textsc{SortEigenelements}}
	\label{algocf:alg_symplectic_sort}
	\begin{algorithmic}[1]
		\Input
		\State $\boldsymbol{M}\in\SpH(2n)$ hermitian positive definite
		\EndInput
		\Output
		\State $(\boldsymbol{\ell}, \boldsymbol{\ell}^{-1}), 
		\boldsymbol{V_0}$
		\EndOutput
		\Procedure{SortEigenelements}{$\boldsymbol{M}$}
		\State $\boldsymbol{\lambda}, \boldsymbol{V} \eqdef 
		\textsc{Diagonalize}(\boldsymbol{M})$
		\State $n\eqdef \textsc{Len}(\boldsymbol{\lambda})/2$
		\State $\text{order} \eqdef \textsc{ArgSort}(\boldsymbol{\lambda})$
		\State $\text{order} \eqdef 
		\textsc{Concatenate}(\text{order}[n:2n], \text{order}[n-1::-1])$ 
		\State \Comment{We use python indexing here, $\boldsymbol{\lambda} 
			= (\boldsymbol{\ell}, \boldsymbol{\ell}^{-1})$.}
		\State \Return $\boldsymbol{\lambda}[\text{order}]$, 
		$\boldsymbol{V}[:,\text{order}]$.
		\EndProcedure
	\end{algorithmic}
\end{algorithm}

\begin{algorithm}[H]
	\caption{\textsc{CsdSp}}
	\label{algocf:alg_CSD}
	\begin{algorithmic}[1]
		\Input
		\State $\boldsymbol{U}\in\SpH(2n)\cap U(2n)$ 
		\EndInput
		\Output
		\State $\boldsymbol{V}, \boldsymbol{W}, \boldsymbol{\theta}$ as 
		defined in  SI Theorem 8 
		\EndOutput
		\Procedure{CsdSp}{$\boldsymbol{U}$}
		\State $n=\textsc{Len}(\boldsymbol{U})/2$
		\State $\boldsymbol{U'}\eqdef 
		\boldsymbol{P}^\dagger \boldsymbol{UP}$ 
		\Comment{$\boldsymbol{U'}\in U(n,n)\cap U(2n)$}
		\State  $\boldsymbol{Q_1} \eqdef \boldsymbol{U'}[:n,:n]$
		\State $\boldsymbol{Q_2} \eqdef \boldsymbol{U'}[n:,n:]$
		\Comment{SI Eq. (III.40)}
		\State $\boldsymbol{W}^\dagger, e^{2\ii \boldsymbol{\theta}} \eqdef 
		\textsc{Diagonalize}(\boldsymbol{Q_2^\dagger Q_1})$ 
		\Comment{SI Eq. (III.41)}
		\State $\boldsymbol{\theta} \eqdef \textsc{Angle}(e^{2\ii 
			\boldsymbol{\theta}})/2$
		\State \Return $ \boldsymbol{QW^\dagger}e^{-\ii 
			\boldsymbol{\theta}}, \boldsymbol{W}, \boldsymbol{\theta}$
		\EndProcedure
	\end{algorithmic}
\end{algorithm}

\begin{algorithm}[H]
	\caption{\textsc{SymplecticSpectral}
	}
	\label{algocf:alg_symplectic_HPD}
	\begin{algorithmic}[1]
		\Input
		\State $\boldsymbol{M}\in\SpH(2n)$ hermitian positive definite
		\EndInput
		
		\Output
		\State $(\boldsymbol{\tilde{D}}, \boldsymbol{\tilde{D}}^{-1})$, 
		$\boldsymbol{U}$ as defined in SI 
		Eq. (III.21)
		\EndOutput
		\Procedure{SymplecticSpectral}{$\boldsymbol{M}$}
		\State $\boldsymbol{\lambda}, \boldsymbol{V} \eqdef 
		\textsc{SortEigenelements}(\boldsymbol{M})$
		\State $n\eqdef \textsc{Len}(\boldsymbol{\lambda})/2$
		\State $\boldsymbol{V_1}\eqdef \textsc{Zeros\_like}(\boldsymbol{V})$
		\State $m' = 0$ \Comment{Half multiplicity of the eigenvalue 1}
		\For{$j\in\textsc{Range}(n)$}
		\If{$\boldsymbol{\lambda}[j] = 1$}  \Comment{SI Eq. (III.2)}
		\State $\boldsymbol{V_1}[:,j]\eqdef\boldsymbol{V}[:,j]$
		\State $\boldsymbol{V_1}[:,n+j]\eqdef 
		-\boldsymbol{J}\boldsymbol{V}[:,j]$
		\Else
		\State $\boldsymbol{V_1}[:,j]\eqdef\boldsymbol{V}[:,j]$
		\State $\boldsymbol{V_1}[:,n+j]\eqdef 
		\boldsymbol{V}[:,n+j]$
		\State $m' \eqdef m'+1$
		\EndIf
		\EndFor \Comment{SI Eq. (III.25)}
		\State $\boldsymbol{\tilde{V}} \eqdef \boldsymbol{V_1}[:, 
		\textsc{Concatenate}(\textsc{Range}(m'), \textsc{Range}(n,n+m'))]$ 
		\Comment{Prepare for SI Eq. (III.26)}
		\State $\boldsymbol{J'}\eqdef \boldsymbol{\tilde{V}}^\dagger 
		\boldsymbol{J} \boldsymbol{\tilde{V}}$ \Comment{$2m' \times 2m'$ 
			matrix}
		\State $\boldsymbol{\lambda'}, \boldsymbol{V'} \eqdef 
		\textsc{Diagonalize}(\ii \boldsymbol{J'})$ \Comment{SI 
		Eq. (III.30)}
		\State $\boldsymbol{\lambda'} = 
		\boldsymbol{\lambda'}[\textsc{Argsort}(- \boldsymbol{\lambda'})]$ 
		\Comment{$\boldsymbol{\lambda'}\simeq \boldsymbol{I_{m'}}$}
		\State $\boldsymbol{V'} = \boldsymbol{V'}[\textsc{Argsort}(- 
		\boldsymbol{\lambda'})]$
		\State $\boldsymbol{W'}\eqdef \boldsymbol{V'P}^\dagger$
		\State $\boldsymbol{W}\eqdef \textsc{Block\_diag}([\boldsymbol{W'}, 
		\boldsymbol{1_{2n-2m'}}])$
		\State $\text{perm\_list} = 
		\textsc{Concatenate}(\textsc{Range}(m'), 
		\textsc{Range}(2m',n+m'), 
		\textsc{Range}(n+m',n+2m'), \textsc{Range}(n+2m',2n))$
		\Comment{$\boldsymbol{Q}^\dagger$ SI 
			Eq. (III.28)}
		\State $\boldsymbol{W} \eqdef 
		\boldsymbol{W}[:,\text{perm\_list}]$\Comment{$\boldsymbol{WQ}^\dagger$}
		\State $\boldsymbol{W} \eqdef \boldsymbol{W}[\text{perm\_list},:]$ 
		\Comment{$\boldsymbol{QWQ^\dagger}$}
		\State $\boldsymbol{U}\eqdef \boldsymbol{V_1W}$
		\State \Return $\boldsymbol{\lambda}, \boldsymbol{U}$  
		\Comment{SI 
			Theorem 7 ($\boldsymbol{\lambda} = 
			(\boldsymbol{\tilde{D}}, \boldsymbol{\tilde{D}}^{-1})$)}
		\EndProcedure
	\end{algorithmic}
\end{algorithm}

\begin{algorithm}[H]
	\caption{ \textsc{SvdSp}
	}
	\label{algocf:alg_bloch_messiah}
	\begin{algorithmic}[1]
		\Input
		\State $\boldsymbol{M}\in\SpH(2n)$ 
		\EndInput
		\Output
		\State $\boldsymbol{U}, \boldsymbol{V}, (\boldsymbol{\tilde{D}}, 
		\boldsymbol{\tilde{D}}^{-1})$ as defined in SI Theorem 6 
		\EndOutput
		
		\Procedure{SvdSp}{$\boldsymbol{M}$}
		\State $\boldsymbol{H}, \boldsymbol{U}\eqdef 
		\textsc{Polar}(\boldsymbol{M})$ \Comment{$\boldsymbol{M} = 
			\boldsymbol{UH}$}
		\State  $\boldsymbol{\lambda}, \boldsymbol{V} \eqdef 
		\textsc{SymplecticSpectral}(\boldsymbol{H})$ 
		\Comment{$\boldsymbol{H} = 
			\boldsymbol{V}\text{diag}(\boldsymbol{\lambda})\boldsymbol{V}^\dagger$}
		\State \Return $\boldsymbol{UV},  
		\boldsymbol{V}^\dagger, \boldsymbol{\lambda}$ 
		\Comment{$\boldsymbol{M} = 
			\boldsymbol{UV}\text{diag}(\boldsymbol{\lambda})\boldsymbol{V}^\dagger$}
		\EndProcedure
	\end{algorithmic}
\end{algorithm}

	\bibliography{refs_quantum_lti}
	
\end{document}